\documentclass[twocolumn]{aastex631}
\usepackage{verbatim}
\usepackage{tabularx}
\usepackage{amsmath}
\usepackage{braket}
\usepackage{booktabs}
\usepackage{graphicx}
\usepackage{amssymb}
\usepackage{bm}	
\usepackage{xcolor}
\usepackage{hyperref}
\newcommand{\prepcite}[1]{\textcolor{blue}{#1}}
\newcommand{\kms}{\,km\,s$^{-1}$}
\newcommand{\mgii}{Mg\hspace{0.5mm}{\footnotesize II}}
\newcommand{\civ}{C\hspace{0.5mm}{\footnotesize IV}}
\newcommand{\Tmgii}{$T\sim \rm 10^{4-4.5}\, K$ }
\newcommand{\Tciv}{$T\sim \rm 10^{5-5.5}\, K$ }
\newcommand{\civewGr}[1]{$EW_{\rm 1548}\,>\,#1\,$\AA}
\newcommand{\ewciv}{$EW_{\rm 1548}$\,}

\newcommand{\modify}[1]{\textcolor{black}{#1}}
\shortauthors{Anand et al.}

\begin{document}

\title{The Cosmic Evolution of C~IV Absorbers at $1.4<z<4.5$\,: Insights from $100,000$ Systems in DESI Quasars}

\shorttitle{CIV absorbers in DESI Quasars}

% Author list file generated with: mkauthlist 1.3.0+44.gbe076f7 
% mkauthlist -f --sort-builder --orcid -j apj DESI-2024-0508_author_list.csv desi-508-authors.tex 
%% Orcid numbers may need \usepackage{orcidlink}.
%% Use \input to call the file

\author[0000-0003-2923-1585]{Abhijeet~Anand}
\affiliation{Lawrence Berkeley National Laboratory, 1 Cyclotron Road, Berkeley, CA 94720, USA}

\author{J.~Aguilar}
\affiliation{Lawrence Berkeley National Laboratory, 1 Cyclotron Road, Berkeley, CA 94720, USA}

\author[0000-0001-6098-7247]{S.~Ahlen}
\affiliation{Physics Dept., Boston University, 590 Commonwealth Avenue, Boston, MA 02215, USA}

\author[0000-0001-9712-0006]{D.~Bianchi}
\affiliation{Dipartimento di Fisica ``Aldo Pontremoli'', Universit\`a degli Studi di Milano, Via Celoria 16, I-20133 Milano, Italy}
\affiliation{INAF-Osservatorio Astronomico di Brera, Via Brera 28, 20122 Milano, Italy}

\author[0000-0002-8934-0954]{A.~Brodzeller}
\affiliation{Lawrence Berkeley National Laboratory, 1 Cyclotron Road, Berkeley, CA 94720, USA}

\author{D.~Brooks}
\affiliation{Department of Physics \& Astronomy, University College London, Gower Street, London, WC1E 6BT, UK}

\author{R.~Canning}
\affiliation{Institute of Cosmology and Gravitation, University of Portsmouth, Dennis Sciama Building, Portsmouth, PO1 3FX, UK}

\author{T.~Claybaugh}
\affiliation{Lawrence Berkeley National Laboratory, 1 Cyclotron Road, Berkeley, CA 94720, USA}

\author[0000-0002-2169-0595]{A.~Cuceu}
\affiliation{Lawrence Berkeley National Laboratory, 1 Cyclotron Road, Berkeley, CA 94720, USA}

\author[0000-0002-1769-1640]{A.~de la Macorra}
\affiliation{Instituto de F\'{\i}sica, Universidad Nacional Aut\'{o}noma de M\'{e}xico,  Circuito de la Investigaci\'{o}n Cient\'{\i}fica, Ciudad Universitaria, Cd. de M\'{e}xico  C.~P.~04510,  M\'{e}xico}

\author{P.~Doel}
\affiliation{Department of Physics \& Astronomy, University College London, Gower Street, London, WC1E 6BT, UK}

\author[0000-0003-4992-7854]{S.~Ferraro}
\affiliation{Lawrence Berkeley National Laboratory, 1 Cyclotron Road, Berkeley, CA 94720, USA}
\affiliation{University of California, Berkeley, 110 Sproul Hall \#5800 Berkeley, CA 94720, USA}

\author[0000-0002-3033-7312]{A.~Font-Ribera}
\affiliation{Institut de F\'{i}sica d’Altes Energies (IFAE), The Barcelona Institute of Science and Technology, Edifici Cn, Campus UAB, 08193, Bellaterra (Barcelona), Spain}

\author[0000-0002-2890-3725]{J.~E.~Forero-Romero}
\affiliation{Departamento de F\'isica, Universidad de los Andes, Cra. 1 No. 18A-10, Edificio Ip, CP 111711, Bogot\'a, Colombia}
\affiliation{Observatorio Astron\'omico, Universidad de los Andes, Cra. 1 No. 18A-10, Edificio H, CP 111711 Bogot\'a, Colombia}

\author{E.~Gaztañaga}
\affiliation{Institut d'Estudis Espacials de Catalunya (IEEC), c/ Esteve Terradas 1, Edifici RDIT, Campus PMT-UPC, 08860 Castelldefels, Spain}
\affiliation{Institute of Cosmology and Gravitation, University of Portsmouth, Dennis Sciama Building, Portsmouth, PO1 3FX, UK}
\affiliation{Institute of Space Sciences, ICE-CSIC, Campus UAB, Carrer de Can Magrans s/n, 08913 Bellaterra, Barcelona, Spain}

\author[0000-0003-3142-233X]{S.~Gontcho A Gontcho}
\affiliation{Lawrence Berkeley National Laboratory, 1 Cyclotron Road, Berkeley, CA 94720, USA}

\author{G.~Gutierrez}
\affiliation{Fermi National Accelerator Laboratory, PO Box 500, Batavia, IL 60510, USA}

\author[0000-0001-9822-6793]{J.~Guy}
\affiliation{Lawrence Berkeley National Laboratory, 1 Cyclotron Road, Berkeley, CA 94720, USA}

\author[0000-0002-9136-9609]{H.~K.~Herrera-Alcantar}
\affiliation{Institut d'Astrophysique de Paris. 98 bis boulevard Arago. 75014 Paris, France}
\affiliation{IRFU, CEA, Universit\'{e} Paris-Saclay, F-91191 Gif-sur-Yvette, France}

\author[0000-0002-6024-466X]{M.~Ishak}
\affiliation{Department of Physics, The University of Texas at Dallas, 800 W. Campbell Rd., Richardson, TX 75080, USA}

\author[0000-0002-0000-2394]{S.~Juneau}
\affiliation{NSF NOIRLab, 950 N. Cherry Ave., Tucson, AZ 85719, USA}

\author{R.~Kehoe}
\affiliation{Department of Physics, Southern Methodist University, 3215 Daniel Avenue, Dallas, TX 75275, USA}

\author[0000-0001-6356-7424]{A.~Kremin}
\affiliation{Lawrence Berkeley National Laboratory, 1 Cyclotron Road, Berkeley, CA 94720, USA}

\author[0000-0003-1838-8528]{M.~Landriau}
\affiliation{Lawrence Berkeley National Laboratory, 1 Cyclotron Road, Berkeley, CA 94720, USA}

\author[0000-0001-7178-8868]{L.~Le~Guillou}
\affiliation{Sorbonne Universit\'{e}, CNRS/IN2P3, Laboratoire de Physique Nucl\'{e}aire et de Hautes Energies (LPNHE), FR-75005 Paris, France}

\author[0000-0003-1887-1018]{M.~E.~Levi}
\affiliation{Lawrence Berkeley National Laboratory, 1 Cyclotron Road, Berkeley, CA 94720, USA}

\author[0000-0003-4962-8934]{M.~Manera}
\affiliation{Departament de F\'{i}sica, Serra H\'{u}nter, Universitat Aut\`{o}noma de Barcelona, 08193 Bellaterra (Barcelona), Spain}
\affiliation{Institut de F\'{i}sica d’Altes Energies (IFAE), The Barcelona Institute of Science and Technology, Edifici Cn, Campus UAB, 08193, Bellaterra (Barcelona), Spain}

\author[0000-0002-1125-7384]{A.~Meisner}
\affiliation{NSF NOIRLab, 950 N. Cherry Ave., Tucson, AZ 85719, USA}

\author{R.~Miquel}
\affiliation{Instituci\'{o} Catalana de Recerca i Estudis Avan\c{c}ats, Passeig de Llu\'{\i}s Companys, 23, 08010 Barcelona, Spain}
\affiliation{Institut de F\'{i}sica d’Altes Energies (IFAE), The Barcelona Institute of Science and Technology, Edifici Cn, Campus UAB, 08193, Bellaterra (Barcelona), Spain}

\author[0000-0002-2733-4559]{J.~Moustakas}
\affiliation{Department of Physics and Astronomy, Siena College, 515 Loudon Road, Loudonville, NY 12211, USA}

\author{A.~Muñoz-Gutiérrez}
\affiliation{Instituto de F\'{\i}sica, Universidad Nacional Aut\'{o}noma de M\'{e}xico,  Circuito de la Investigaci\'{o}n Cient\'{\i}fica, Ciudad Universitaria, Cd. de M\'{e}xico  C.~P.~04510,  M\'{e}xico}

\author[0000-0002-5166-8671]{L.~Napolitano}
\affiliation{Department of Physics \& Astronomy, University  of Wyoming, 1000 E. University, Dept.~3905, Laramie, WY 82071, USA}

\author[0000-0001-6979-0125]{I.~P\'erez-R\`afols}
\affiliation{Departament de F\'isica, EEBE, Universitat Polit\`ecnica de Catalunya, c/Eduard Maristany 10, 08930 Barcelona, Spain}

\author{G.~Rossi}
\affiliation{Department of Physics and Astronomy, Sejong University, 209 Neungdong-ro, Gwangjin-gu, Seoul 05006, Republic of Korea}

\author[0000-0002-9646-8198]{E.~Sanchez}
\affiliation{CIEMAT, Avenida Complutense 40, E-28040 Madrid, Spain}

\author{D.~Schlegel}
\affiliation{Lawrence Berkeley National Laboratory, 1 Cyclotron Road, Berkeley, CA 94720, USA}

\author{M.~Schubnell}
\affiliation{Department of Physics, University of Michigan, 450 Church Street, Ann Arbor, MI 48109, USA}
\affiliation{University of Michigan, 500 S. State Street, Ann Arbor, MI 48109, USA}

\author{D.~Sprayberry}
\affiliation{NSF NOIRLab, 950 N. Cherry Ave., Tucson, AZ 85719, USA}

\author[0000-0003-1704-0781]{G.~Tarl\'{e}}
\affiliation{University of Michigan, 500 S. State Street, Ann Arbor, MI 48109, USA}

\author[0000-0001-8433-550X]{M.~J.~Temple}
\affiliation{Centre for Extragalactic Astronomy, Department of Physics, Durham University, South Road, Durham, DH1 3LE, UK}

\author{B.~A.~Weaver}
\affiliation{NSF NOIRLab, 950 N. Cherry Ave., Tucson, AZ 85719, USA}

\author[0000-0001-5381-4372]{R.~Zhou}
\affiliation{Lawrence Berkeley National Laboratory, 1 Cyclotron Road, Berkeley, CA 94720, USA}

\correspondingauthor{Abhijeet~Anand}
\email{AbhijeetAnand@lbl.gov}

\begin{abstract}

We present the largest catalog to date of triply ionized carbon (C~\textsc{iv}) absorbers detected in quasar spectra from the Dark Energy Spectroscopic Instrument. Using an automated matched-kernel convolution method with adaptive signal-to-noise thresholds, we identify $101,487$ \civ~systems in the redshift range \( 1.4 < z < 4.5 \) from $300,637$ quasar spectra. Completeness is estimated via Monte Carlo simulations and catalog is $50\%$ complete at \(\mathrm{EW}_{\mathrm{CIV}} \geq 0.4\)~\AA.~ The differential equivalent width frequency distribution declines exponentially and shows weak redshift evolution. The absorber incidence per unit comoving path increases by a factor of $2-5$ from \( z \approx 4.5 \) to \( z \approx 1.4 \), with stronger redshift evolution for strong systems. Using column densities derived from the apparent optical depth method, we constrain the cosmic mass density of C~\textsc{iv}, \(\Omega_{\mathrm{CIV}}\), which increases by a factor of \modify{$\sim 3.8$} from \modify{\( (0.82 \pm 0.05) \times 10^{-8} \)} at \( z \approx 4.5 \) to \modify{\( (3.16 \pm 0.2) \times 10^{-8} \)} at \( z \approx 1.4 \). From \( \Omega_{\rm CIV} \), we estimate a lower limit on intergalactic medium metallicity \( \log(Z_{\rm IGM}/Z_{\odot}) \gtrsim -3.25 \) at \( z \sim 2.3 \), with a smooth decline at higher redshifts. These trends trace the cosmic star formation history and He~\textsc{ii} photoheating rate, suggesting a link between \civ~enrichment, star formation, and UV background over $\sim 3$ Gyr. The catalog also provides a critical resource for future studies connecting circumgalactic metals to galaxy evolution, especially near cosmic noon.

\end{abstract}

\keywords{Quasar absorption line spectroscopy (1317); Intergalactic medium (813); Redshift surveys(1378), Astronomy software(1855)}

%%%%%%%%%%%%%%%%% INTRODUCTION %%%%%%%%%%%%%%%%%%

\section{Introduction}\label{intro}

The interplay between gas inflows and outflows—known as the ``cosmic baryon cycle"—within the galactic disc, halo, and their interaction with the intergalactic medium (IGM) is fundamental to galaxy formation and evolution. Metals synthesized in stars are expelled into the interstellar medium (ISM) or circumgalactic medium (CGM) through mechanisms such as stellar winds, supernovae, and active galactic nuclei (AGN) feedback \cite[see][for reviews]{keres05, somerville15,peroux20a, fumagalli2024}. These metals enrich the CGM and IGM, forming a reservoir from which galaxies can accrete cold gas over time \citep{angles17,nelson20}. This cyclical exchange of gas and metals plays a crucial role in galaxy evolution \citep{oppenheimer08,muratov17}, making it essential to trace the distribution and evolution of metals within the ISM, CGM, and IGM to fully understand the metal and baryon cycles that shape the universe.

Due to the low density and diffuse nature of metal-enriched gas in the CGM and IGM, detecting this gas in emission is challenging, as \modify{emissivity scales with the square of the gas density} and the flux diminishes with distance. In contrast, quasar absorption line spectroscopy provides a direct and powerful method to study gas-phase metals in the CGM and IGM across a broad range of redshifts. \modify{Since absorption strength scales linearly with gas density, even the diffuse CGM can be readily detected in absorption.} By observing metal absorption against the bright light of background quasars, this technique allows us to probe the properties of multiphase gas. Since the strength of absorption lines is not influenced by redshift, this method is particularly effective for constraining the properties of gas over cosmic time.

Among the most extensively studied metal absorption lines are the \mgii\ and \civ\ doublets, which trace different gas phases in the CGM or IGM. The \mgii~$\lambda\lambda 2796, 2803$ doublet is a tracer of low-ionization, cool gas (\Tmgii) in the CGM, intracluster medium (ICM) and IGM \citep{zhu13a,zhou2023LRG, anand2021, anand2022, zhou2023LRG, napolitano2023}, while the \civ~$\lambda\lambda 1548, 1550$ doublet traces warm ionized gas (\Tciv) in these regions \citep{landoni16, tumlinson17, davies2021,schroetter21}. 

Carbon, the fourth most abundant metal in the Universe, plays a crucial role in the chemical evolution of galaxies \citep{henning1998}, and the \civ\ doublet’s relatively high oscillator strength makes it easily observable in the optical/NIR at redshifts $1.4 < z < 4.5$, both in absorption \citep{sargent88, petitjean94, songaila01, cooksey13,hasan20,hasan21,davies2021,monadi2023} and emission \citep{arrigoni15a, arrigoni15b, piacitelli22}. The distinctive doublet nature of both \civ~and \mgii\ also enables automated searches in quasar spectra \citep{chensdss18, anand2021, napolitano2023}, facilitating large-scale studies of metal distribution in the Universe.

Thanks to both low- and high-resolution spectroscopic surveys, our understanding of the cosmic enrichment cycle has improved by investigating \civ~absorber properties detected in the background quasars. Previous analyses using thousands of \civ\ systems have revealed that the equivalent width (or column density) frequency distributions typically follow exponential or power-law shapes, with slopes ranging from $\alpha \approx -1.8$ to $-2.7$ \citep{songaila01, dodorico10, simcoe11, boksenberg15}. Furthermore, the number of \civ\ absorbers per unit redshift, $dN/dz$, increases smoothly from $z=6$ to $z=0$ for strong absorbers (\civewGr{1}), suggesting a significant increase in the abundance of carbon in the late universe compared to earlier epochs \citep{peroux04,cooksey13}.

Using the intermediate strength \civ~absorbers (\civewGr{0.6}) detected in low-resolution quasar spectra, previous studies have revealed that \civ\ mass density, $\Omega_{\rm \civ}$, increases with decreasing redshift, rising from $z=5$ to $z=0$ \citep{songaila01, simcoe11, cooksey13,shull2014}. Similarly, the comoving \civ\ line density, $dN/dX$, grows from $z=5.5$ to $z=0$, though $\Omega_{\rm \civ}$ increases more steeply than $dN/dX$ \citep{cooksey13, codoreanu18}. The ratio of the doublet lines ($EW_{\rm 1548}/EW_{\rm 1550}$) of \civ\ absorbers has also been found to decrease over time, indicating a rise in the mean \civ\ density as the universe evolves \citep{peroux04, songaila05}.

Constraining the cosmic mass density of \civ~absorbers in the Universe across a wide range of redshifts using large datasets from cosmological surveys is essential. In this paper, we study the statistical evolution of different populations of \civ\ absorbers using the latest dataset from the ongoing Dark Energy Survey Instrument (DESI) experiment. Our analysis represents the largest statistical study of quasars and \civ\ absorbers to date, providing significantly reduced measurement errors and allowing us to explore properties in finer redshift and equivalent width (EW) bins than in previous studies.

We organize this paper as follows: Section~\ref{desidata} summarizes the DESI survey and quasar sample. We describe the quasar continuum fitting and absorber detection algorithm in Section~\ref{methods}. In Section~\ref{civ_catalog}, we present the catalog, analyze the \modify{average} properties of individual absorbers\footnote{\modify{Absorption complexes often resolve into several "cloudlets" (or individual absorbers) in higher-resolution data with velocity widths of a few tens of \kms}.}, and discuss the catalog completeness. Section~\ref{results} presents a statistical evolution of \civ~absorbers as a function of redshift and equivalent widths. Finally, we compare our findings with previous studies and discuss the implications of our results in Section~\ref{discussion}. Throughout this paper, we assume a Planck cosmology \citep{planck16} with $\Omega_{\rm m} = 0.307$, $H_0 = 67.7$ \kms\,$\rm Mpc^{-1}$, and $\Omega_{\rm \Lambda} = 1 - \Omega_{\rm m}$.

%%%%%%%%%%%%%%%%% DATA %%%%%%%%%%%%%%%%%%

\section{DESI Survey}\label{desidata}

DESI is a Stage-IV cosmological survey designed to collect millions of spectra of galaxies, stars, and quasars. The primary goal of the DESI mission is to study the large-scale structure and matter clustering in the Universe and constrain cosmological parameters with unprecedented accuracy \citep{DESI2016-Overview, DESI2016-Instrument} by mapping the distribution of quasars and galaxies over a wide range of redshifts.

The DESI instrument is a multi-object spectrograph installed on the Mayall 4-meter telescope at Kitt Peak National Observatory (KPNO) in Arizona. The focal plane is designed to simultaneously point at $5000$ locations on the sky using $5000$ robotic positioners with optical fibers \citep{desifiber2024}, which direct light from these objects to $10$ spectrographs \citep{DESI2016-Instrument, DESI2022-Instrument}. Finally, each spectrum is reduced on a uniform wavelength grid spanning \( 3600 \)~\AA~ to \( 9800 \)~\AA, with a pixel size of \( 0.8 \)~\AA. Survey targets are selected from imaging data obtained with the DESI Legacy Imaging Survey \citep{dey2019}. DESI will collect over 5 million quasar spectra at $0.4<z<5$ by the end of $2026$ that are particularly suited for absorption line studies and can provide large absorber catalogs. Quasars are key tracers of large-scale structures at $z>1.6$. Their detailed target selection process and redshift measurements are discussed in \citet{chaussidon2023}. The first set of data, collected during the survey validation and commissioning phase, was released as part of the Early Data Release (EDR) in June $2023$ \citep{desiedr2023, desisv2023}. 

Here, we use data from DESI Data Release 1 \citep{desidr1release2025} and 2 (DR1 \& DR2)\footnote{DR1 was publicly released on March 19, 2025 \citep{desidr1release2025}, while DR2 will be publicly released the following year.}. DR1 is the first major DESI data release\footnote{\url{https://data.desi.lbl.gov/doc/releases/dr1/}}, superseding the EDR. It includes approximately $14.5$ million extragalactic spectra and $4$ million stellar spectra collected during the survey’s first year \citep{desidr1release2025}. DR2 is a superset of DR1, containing over $33$ million extragalactic and $12$ million stellar spectra from the first three years of observations. All data products were processed using the latest version of \texttt{desispec}, which incorporates improved calibrations, sky subtraction, and spectral modeling \citep{guy2023}.

\begin{figure}
    \centering
    \includegraphics[width=0.975\linewidth]{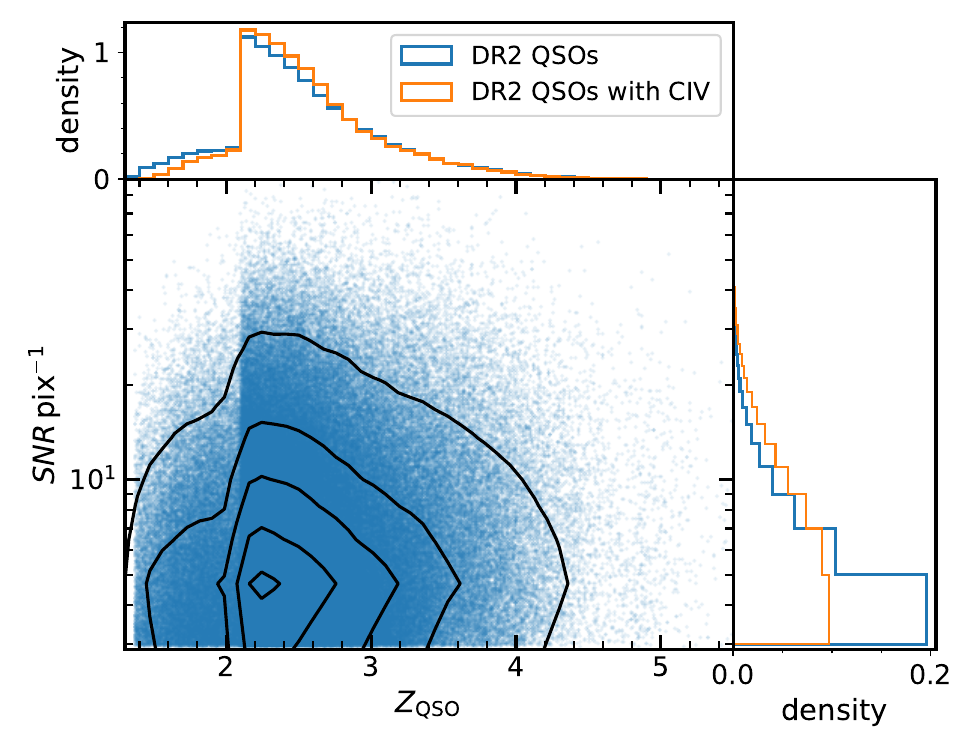}
    \caption{Average signal-to-noise vs. redshift of DR2 quasar sample used for \civ~absorber search in this work. The blue curve is for the parent quasar sample, while the orange represents the quasars with \civ~absorbers. The sharp increase in number density around $z\sim 2.1$ is due to repeated observations of quasars targeting the Ly$\alpha$ forest. The black contour lines indicate the 5th, 25th, 50th, 75th, and 95th percentiles of the distribution.} 
    \label{fig:snr-z-qso}
\end{figure}

\subsection{Quasar Data}

In this work, we use quasars from both \modify{DR1} and DR2 \modify{(DR1 is a subset of DR2)} to search for \civ~absorbers. However, we present results (in section~\ref{results}) only for absorbers identified in DR2 quasars, as the sample size is significantly larger than in DR1. DESI quasar targets are selected using a random forest algorithm applied to the $g-r$ vs. $grz-W1$ color space \citep{chaussidon2023}. Quasar redshifts are fitted using a combination of a PCA-template-based method and a convolutional neural network \citep{busca18, anand24, bailey2025}.

We start with the parent cumulative quasar catalog (\( N \sim 2.4 \) million) and apply the redshift (\( z>1.4 \)) and effective exposure time (\( T_{\rm eff} > 1500\,\rm s \), corresponding to $TSNR^{2}>130$)\footnote{\( T_{\rm eff} = 12.15 \times TSNR^{2} \), where \( \rm TSNR^{2} \) is the squared template signal-to-noise ratio for the LRG target class. The normalization is set so that it equals the exposure time under ideal observing conditions. For more details, see \citet{guy2023}.} cuts to select the first sample. The redshift cut ensures that \civ~absorbers fall within the DESI spectrograph's wavelength range, while the effective time cut is applied to ensure that we select targets observed at least once with the survey's nominal exposure time. We then estimate the continuum using a simple mean transmission flux model and a color-slope-based approach, as described in the following section. We summarize the sample statistics for both DR1 and DR2 in Table~\ref{tab:sample_comparison} and compare them with previous statistics from SDSS. 

\begin{table}[h]
    \centering
    \caption{Sample Statistics for different catalogs. Refs: DESI (\prepcite{Current}), SDSS: \citet{cooksey13}.}
    \begin{tabular}{cccc} % 'l' for left-aligned, 'c' for centered columns
        \toprule
        Object & DESI & DESI & SDSS \\
        & (DR2) & (DR1) &(DR7)  \\
        \midrule
        Total Quasars & $787,166$\footnote{Selection cuts: $130 < \mathrm{TSNR^{2}} <400$ and $z\geq1.4$.} & $241,862$ & $48,260$\footnote{for SDSS, $z\geq 1.7$.}\\
        Selected Quasars\footnote{for DESI: $\braket{SNR}\geq 3\, \rm pix^{-1}$ and for SDSS: $\braket{SNR}\geq 4\, \rm pix^{-1}$.} & $300,637$ & $94,986$ & $26,030$ \\
        \civ~absorbers\footnote{In selected quasars.} & $101,487$& $32,321$ & $14,772$\\
        \civ~($EW_{\rm 1548} > 0.4$\,\AA) & $73,010$ & $23,385$ & $13,514$ \\
        \bottomrule
    \end{tabular}
    \label{tab:sample_comparison}
\end{table}

\begin{figure*}
    \centering 
    \includegraphics[width=0.95\linewidth]{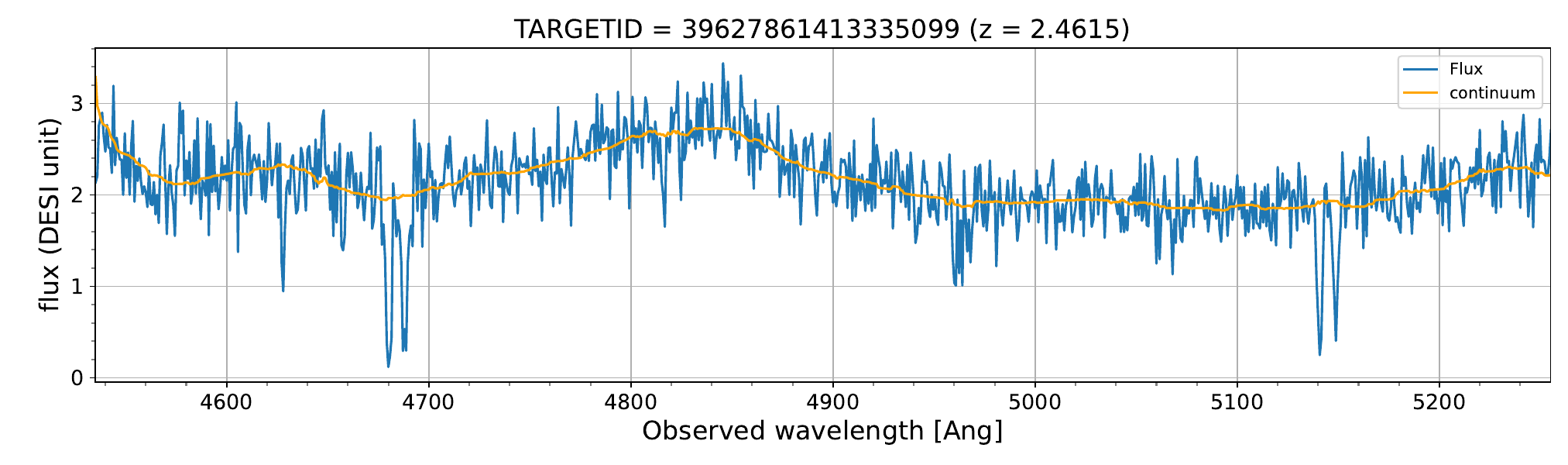}
    \includegraphics[width=0.95\linewidth]{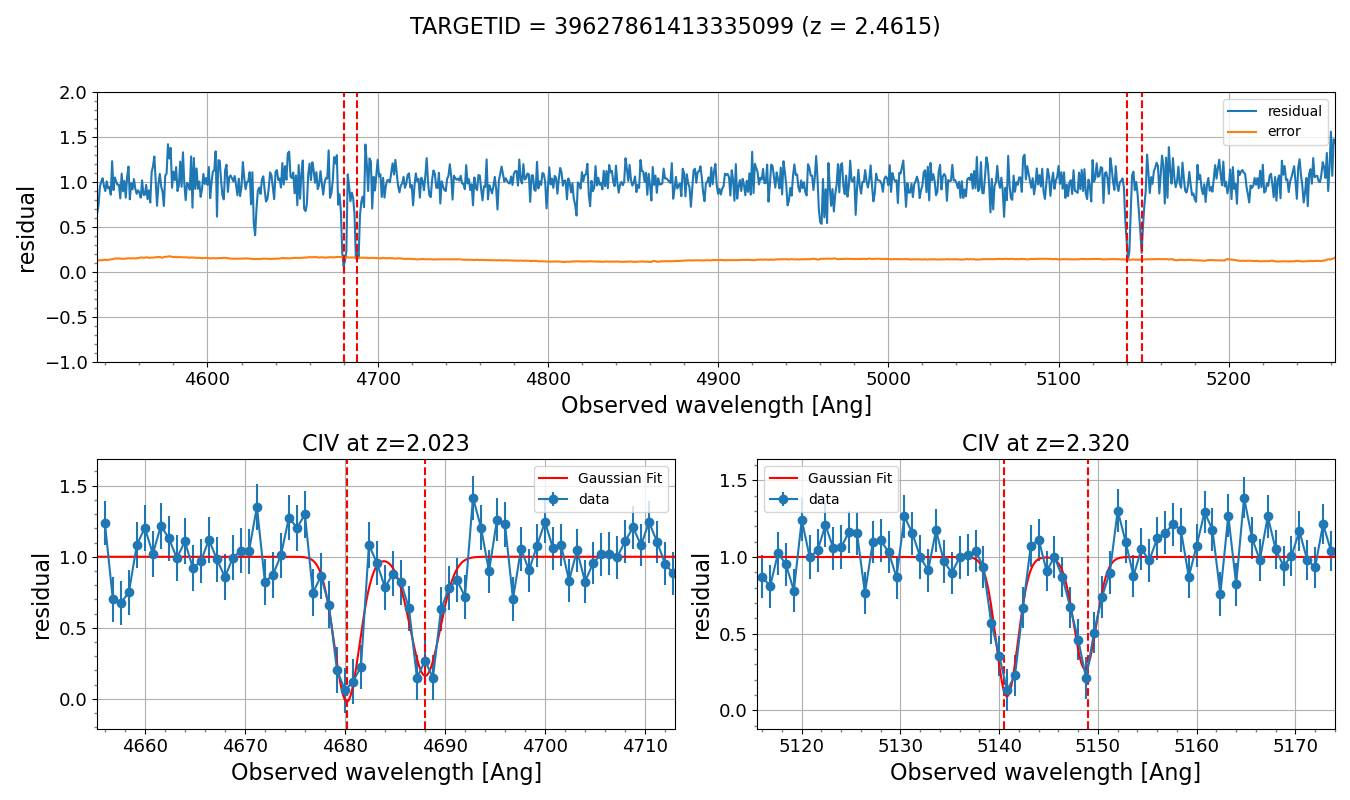}
    \caption{Example DESI spectrum of a quasar with two detected \civ~absorbers at \( z = 2.023 \) and \( z = 2.320 \). \textbf{Top:} Observed DESI flux spectrum (blue) and our continuum estimate (orange).  \textbf{Middle:} Continuum-normalized residual spectrum with the corresponding error.  \textbf{Bottom:} Zoom-in on the absorption features, where the solid red line shows the best-fit double Gaussian used to measure the doublet properties.  The residual spectrum remains mostly flat around \( R \sim 1 \), indicating a fairly well continuum normalization.}
    \label{fig:desispectra}
\end{figure*}

\section{Methods}\label{methods}

\subsection{Quasar Continuum Fitting}

Before detecting any absorber in the quasar spectrum, it is crucial to estimate a robust continuum. The most commonly used empirical method is principal component analysis (PCA). While PCA is mathematically simple and powerful, its eigenvectors do not correspond to any physical properties of the underlying data and can sometimes produce unphysical models, such as negative fluxes \citep[see][for more details]{anand24}. A more recent approach for quasar continuum modeling is nonnegative matrix factorization (NMF) \citep{allen2011,zhu13a, anand2021, napolitano2023}, which factorizes the quasar flux matrix into two smaller nonnegative matrices, ensuring that the modeled flux remains nonnegative by construction. However, this method does not guarantee convergence to a global minimum \citep[see][for more details]{zhu2016} and can give slightly different models depending on the random initial conditions.

In the current study, we exploit another simple approach for modeling the quasar continuum, which has been developed in recent years for cosmological 1D and 3D power spectrum analyses \citep{dumasbourboux2020, karacayli2024}. The algorithm is publicly available\footnote{\url{https://github.com/igmhub/picca}} as part of the Package for IGM Cosmological-Correlations Analyses (\texttt{picca}) and has been used for BAO analysis with DESI data. It models the quasar continuum as the product of the mean transmission, \( \bar{F}(\lambda) \), of the IGM and the unabsorbed quasar continuum, \( C_{\lambda} \), intrinsic to the quasar, where \( \lambda \) is the observed wavelength \citep{dumasbourboux2020}. This product serves as the `average quasar spectrum' in the rest-frame wavelength, \( \bar{C}(\lambda_{\rm rest}) \), corrected by an amplitude (\( a_{q} \)) and a slope term (\( b_{q} \)) in \(\log \lambda\) space. Mathematically, the fitted continuum, \( F_{c, q} (\lambda) \), is expressed as:
\begin{equation}
    F_{c, q} (\lambda) = \bar F(\lambda)C_{q}(\lambda) = \bar C(\lambda_{\rm rest})(a_{q} + b_{q} log\lambda)
\end{equation}
The rest-frame \textit{average quasar spectrum}, \( \bar{C}(\lambda_{\rm rest}) \), can be constructed by stacking multiple quasar spectra in their corresponding rest frames \citep{vandenberk2001}. The linear fitting parameters \( a_{q} \) and \( b_{q} \) are found by maximizing the log-likelihood function. These two linear coefficients basically correct for the variation of brightness and color of the quasar in the wavelength region of interest (CIV forest in our case, see \citealt[][for details]{dumasbourboux2020}). \texttt{picca} allows quasar continuum measurements in a user-defined rest-frame wavelength region. However, fitting this simple combination of the average quasar spectrum and slope correction does not produce an adequate continuum across the entire spectrum, as the method lacks the flexibility to account for intrinsic quasar spectral variations over a broad wavelength region. Therefore, we restrict our continuum estimation to the quasar rest-frame wavelength range: $1310\,\text{\AA} \leq \lambda_{\rm rest} \leq 1520\,\text{\AA}$. This wavelength range is selected such that \civ~absorbers lie between the quasar's intrinsic O~\textsc{i} and C~\textsc{iv} emission lines. The upper limit is set to ensure that the absorber is blueshifted from the background quasar by at least \( \Delta v = -5000 \,\rm km/s \) \citep{cooksey13} \footnote{Although we use this limit, previous studies \citep{bowler14} have suggested that absorbers with velocities as high as \( \delta v = -10000 \,\rm km/s \) could still be associated with quasar outflows.}. After computing the quasar continuum, we define the `residual' spectrum, \( R(\lambda) \), as the ratio of the observed flux to the measured continuum:
\begin{equation}
    R(\lambda) = \frac{Flux}{Continuum} = \frac{f_{q}(\lambda)}{F_{c, q} (\lambda)}
\end{equation}
where $f_{q}$ is the observed-frame quasar flux and $\lambda = (1+z_{\rm qso})\times \lambda_{\rm rest}$ is the observed wavelength. The final error estimate on the residual includes both the pipeline error estimate and a large-scale structure variance term. 

We further smooth the residual using a median filtering approach with a kernel size of $71$ pixels (taken to be $\sim 5-8$ larger than the typical line widths of \civ~systems, \modify{which span $\sim10$--$15$ pixels on the DESI wavelength grid}) to remove intermediate fluctuations. The final residual is obtained by dividing the original residual by the smoothed residual. For consistency, the errors are also normalized by the same smoothed residual. Furthermore, we perform a median stacking of quasar residuals in the rest frame of quasars for our parent sample to assess the quality of the continuum estimation. The composite spectrum is shown in Appendix~\ref{qso_residual} (Figure~\ref{fig:residual_qso_frame}). We find that the residual remains mostly flat ($R\sim1$), with variations of \( \lesssim 1\% \) near the C\,\textsc{ii} and Si\,\textsc{iv} emission features of quasars. This further confirms that our continuum modeling performs pretty well in general.

Finally, for our parent sample used to search for \civ~absorbers, we select only quasars with an average signal-to-noise ratio $3$, i.e., $\braket{SNR}>3\,\rm pix^{-1}$. This selection reduces the sample size by a factor of $\sim 2.5$. The final sample consists of $300,637$ quasars out of $787,166$ in DR2, with a similar reduction observed for the DR1 sample (see Table~\ref{tab:sample_comparison}). The redshift vs. mean signal-to-noise (SNR) distributions of parent DR2 quasars used for \civ~search are shown in  Figure~\ref{fig:snr-z-qso}. The sharp rise at \( z\sim 2.1 \) is due to a higher number of quasar targets that were observed to study the Ly$\alpha$ forest, which is used for Baryon Acoustic Oscillation (BAO) analysis \citep{desicosmo2024,desidr2bao2025, desidr2cosmo2025}. They were observed more than once to achieve higher effective exposure times than the typical value (\( T_{\rm eff} \sim 1000 \,\rm s \)). Most of the quasars have mean $\braket{SNR}<20$ in our sample, with a median value of $\braket{SNR}\sim 5.9$, while the median value of quasars with at least one absorber is $\braket{SNR}\sim 8.7$. We see that the majority of the quasars have average $\braket{SNR}<10$. 

\begin{figure}
    \centering
    \includegraphics[width=0.975\linewidth]{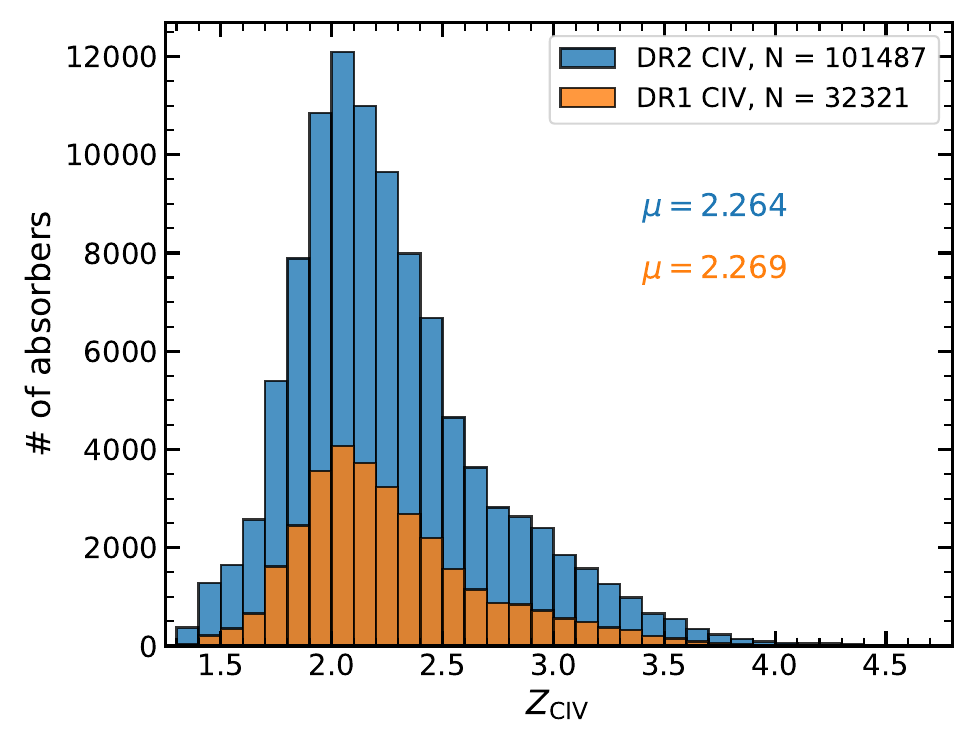}
    \caption{Redshift histogram of \civ~absorbers detected in DR1 (orange) and DR2 (blue) quasars. The mean redshifts are $\braket{z}\sim 2.269$ and $\braket{z}\sim 2.264$, respectively.}
    \label{fig:civ_redshift}
\end{figure}

\subsection{Absorber Detection Algorithm}

The \civ~absorber is a doublet with rest-frame wavelengths $\lambda\lambda 1548, 1550$ \AA, and can, therefore, be detected in the optical at $z\geq1.35$. This doublet nature allows the development of an automated algorithm for their detection in quasar spectra. Also, given the millions of quasar spectra collected by DESI, visual identification is impractical. We use an automated doublet finder, \texttt{qsoabsfind}\footnote{\url{https://github.com/abhi0395/qsoabsfind/releases/tag/v1.0.3}}, a \texttt{Python} module, originally developed by \citet{anand2021} to search for \mgii~absorbers in SDSS DR16 quasars. Due to the similarity in instrument properties, resolution\footnote{The instrumental FWHM resolution of DESI spectrographs varies between \( 60\text{--}150 \,\mathrm{km\,s^{-1}} \), with the blue arm (B-camera) having the lowest resolution and the near-IR arm (Z-camera) having the highest \citep{DESI2022-Instrument, guy2023}.}, wavelength coverage, and typical SNR, we extend this method to DESI quasars \citep{qsoabsfind2025}. The module employs Python's \texttt{multiprocessing} parallelization and can process 1000 spectra in \( \sim 90 \) seconds on a single NERSC\footnote{National Energy Research Scientific Computing Center, \url{www.nersc.gov/}} node with 256 CPU cores. For a detailed description of the algorithm, we refer readers to \citet{anand2021}, and here, we provide only a brief summary of the method.

First, we define the \textit{absorber search window}, the wavelength range within which we search for \civ~absorbers. The minimum and maximum wavelengths are selected using the following criteria:

\begin{equation}
\begin{split}
\lambda_{\rm start} &= \max \{\lambda_{\rm obs, \, min},\, 1310 \cdot(1+z_{\rm qso})\} + \delta_{\lambda} \\
\lambda_{\rm end} &= \min \{\lambda_{\rm obs, \, max},\, 1520 \cdot(1+z_{\rm qso})\} - \delta_{\lambda}
\end{split}
\label{eqn:wavewindow}
\end{equation}

where \(\delta_{\lambda} = 12 \)~\AA, corresponding to $15$ DESI pixels. This slight offset is applied to avoid spectral edges, where pixels can be noisy, and to ensure that both lines fall within the spectrum. $\lambda_{\rm min}=3600$\,\AA~and $\lambda_{\rm max}=9824$\,\AA~are the observed wavelength edges of the DESI spectra. While defining our search window, we also mask the Ca\,\textsc{ii} $\rm \lambda\lambda 3934, 3969$ lines (due to confusion with \civ~absorbers at $z\sim 1.54$) and other skylines (such as $\rm \lambda 5577$ and $\lambda 6300$ and OH lines).

The automatic absorber detection method employs a matched Gaussian kernel convolution approach and user-defined thresholds on adaptive signal-to-noise to detect \civ~doublets. The continuum-normalized spectrum is first convolved at each wavelength pixel using a double Gaussian kernel mimicking the \civ~absorber profile. An adaptive average SNR threshold ($2\sigma$), defined as the standard deviation of the convolved flux within \( \pm 200 \) pixels around each pixel, is then applied to the convolved spectrum to identify potential absorber pixels. Next, we fit a double Gaussian profile with varying amplitudes, line widths, and line centers using standard Levenberg-Marquardt minimization to determine the absorber redshifts and equivalent widths. During the fitting process, we incorporate uncertainties from the residual spectrum to estimate errors on redshift and equivalent width.  

To select the best-fit absorbers, we retain only systems where the line separation is within $1$~\AA~ of the true value\footnote{$\lambda_{1}=1548.20$~\AA,\, $\lambda_{2}=1550.77$~\AA,\, $\Delta \lambda = 2.57$~\AA\, $f_{1} = 0.192, \, f_{2}=0.096$, where $f_{1}$ and $f_{2}$ are oscillator strengths of the transition.} and the doublet ratio (DR) is within $1-\sigma_{DR}<DR<2+\sigma_{DR}$, where uncertainty on doublet ratio ($\sigma_{\rm DR}$) is estimated by propagating equivalent widths errors, which were previously calculated from fit uncertainty. We additionally require that equivalent widths satisfy \( \mathrm{EW} / \sigma_{\mathrm{EW}} > 1 \) for both lines, and the instrumental-resolution-corrected\footnote{We subtract the instrumental resolution from the fitted line widths in quadrature to estimate the intrinsic broadening.} intrinsic line width must also be non-negative. Finally, we compute the local SNR for each line in the doublet \footnote{$S/N = \displaystyle\sum_{i=j_{1}}^{j_{2}}\Big(1-\frac{F_{i}}{F_{C,i}}\Big)\, / \, \Big(\sum_{i=j_{1}}^{j_{2}}\sigma_{i}^{2}\Big)^{1/2}\; \,$ where \( F \) and \( F_{C} \) are the flux and continuum, respectively, and \( \sigma \) is the corresponding error. The indices \( j_{1} \) and \( j_{2} \) define the range of pixels from the line centers where the best-fit residual approaches unity.} and then apply the \civ~\textit{doublet SNR criteria} to finalize our absorber selection. 
\begin{center}
    $\rm S/N (\lambda1548) 
    \geq 3  \text{ and } S/N (\lambda1550) \geq 2$
\end{center}
The final candidate in our catalog is selected if it satisfies the \civ~\textit{doublet SNR criteria}. This selection process, based on physical properties, effectively rejects false positives and has been successfully applied in previous studies \citep{zhu13a,anand2021}.

\begin{figure*}
    \centering
    \includegraphics[width=0.42\linewidth]{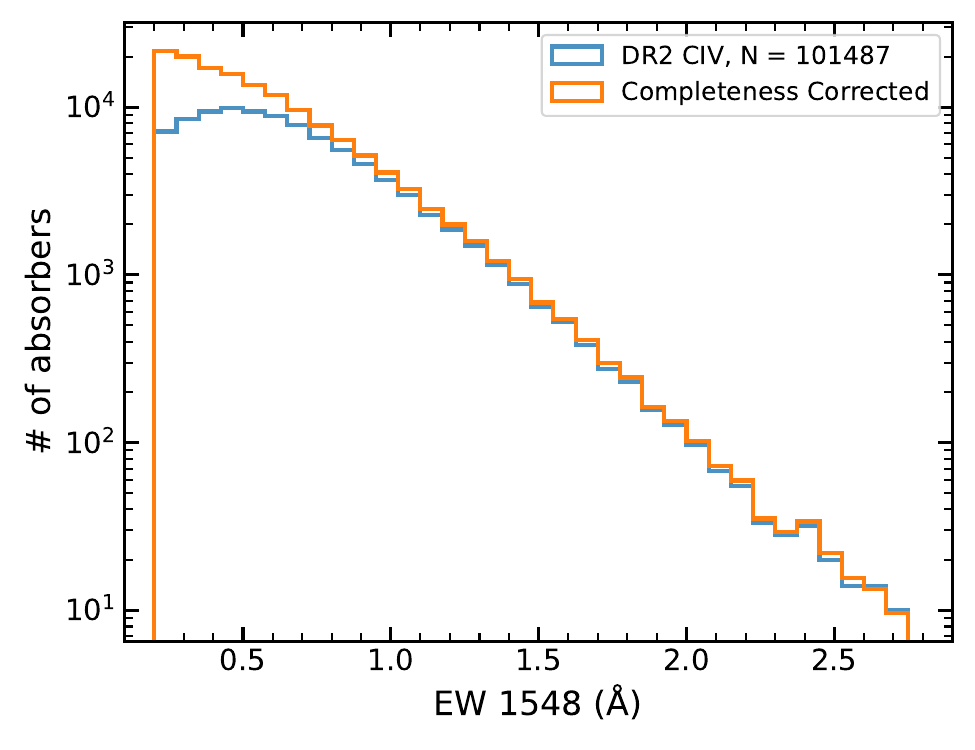}
    \includegraphics[width=0.43\linewidth]{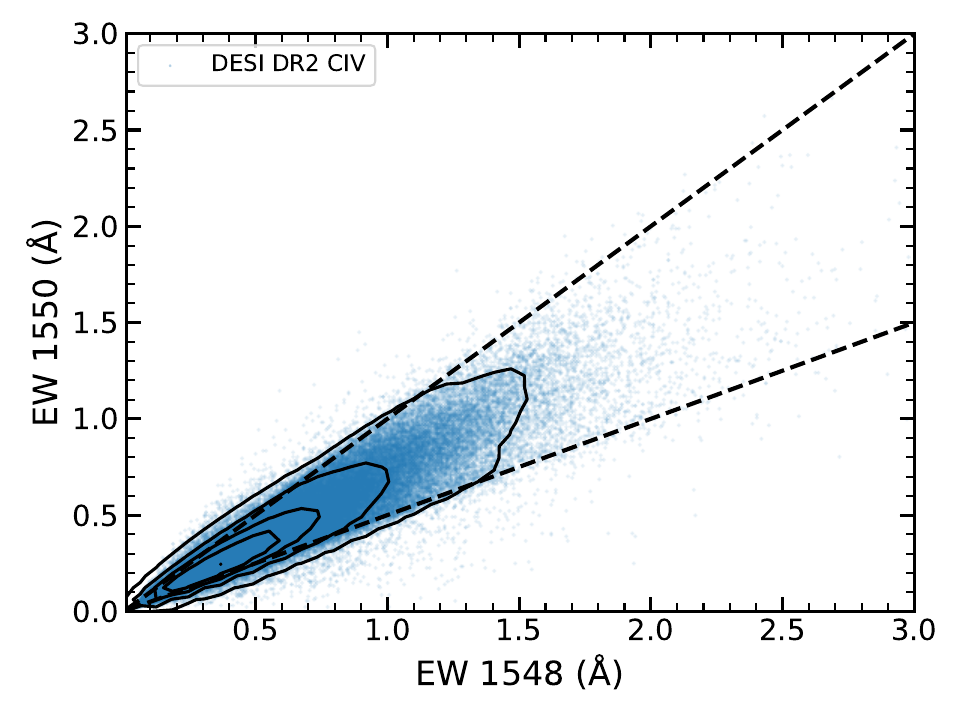}\\
    \includegraphics[width=0.43\linewidth]{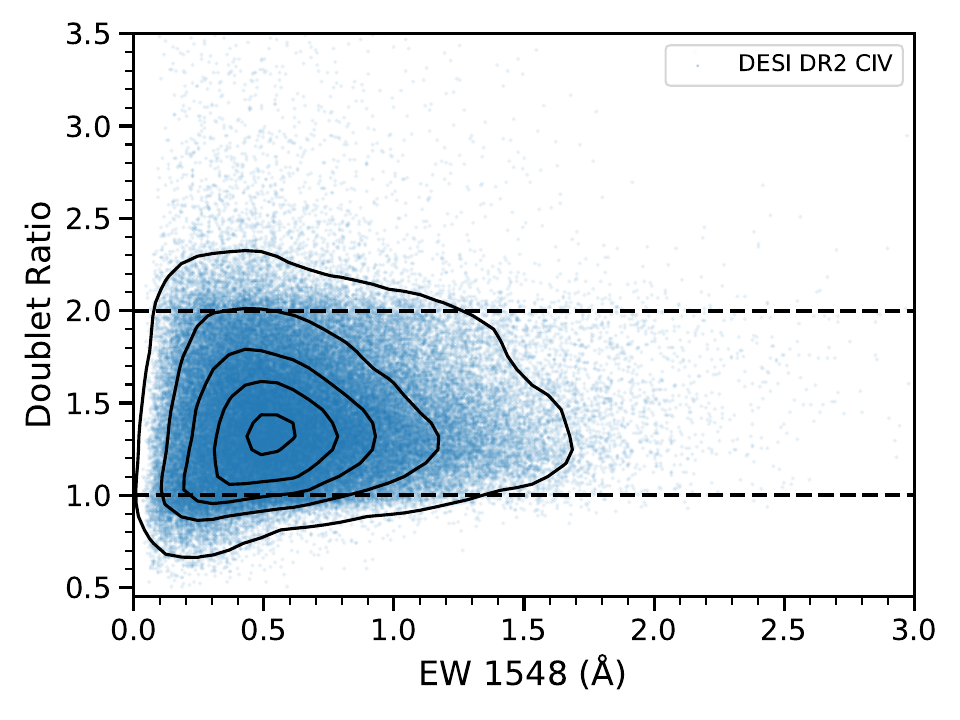}
    \includegraphics[width=0.43\linewidth]{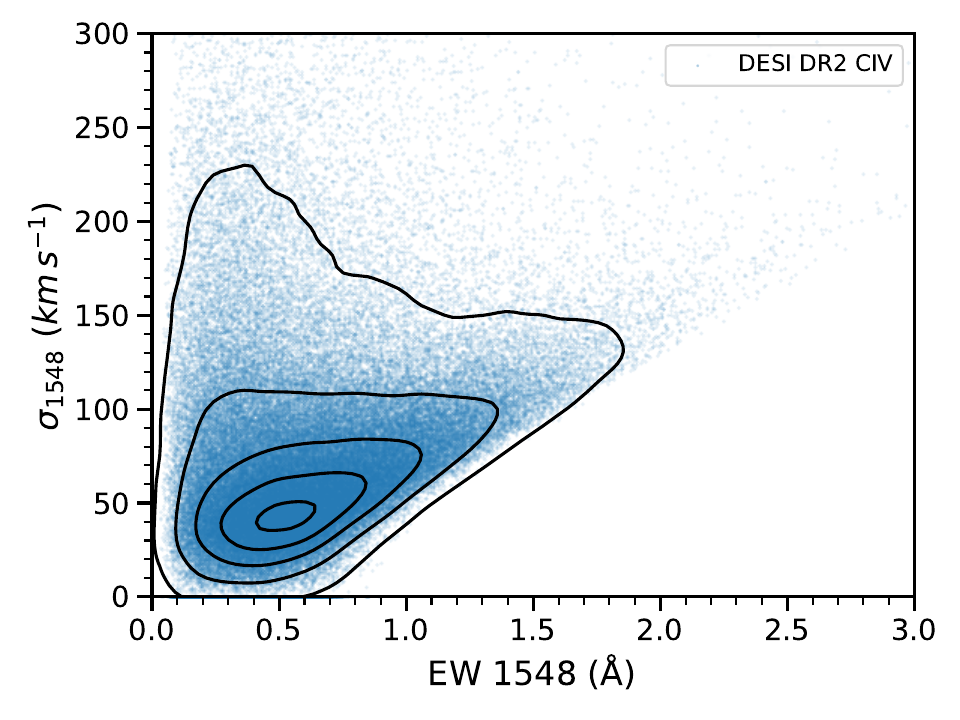}
    \caption{\civ~absorber physical properties. \textbf{Top left:} The observed (blue) and completeness-corrected (orange) distribution of \ewciv. The turnover at $\sim 0.5$~\AA\, in the observed distribution, is not real and is due to low completeness, as seen in the corrected distribution. \textbf{Top right: }Rest-frame equivalent width comparison for both lines of \civ~doublets. The black dashed lines show the $y=x$ and $y=x/2$ curves, derived from the oscillator strengths of the lines. \textbf{Bottom left:} Doublet ratio vs \ewciv for detected systems. The horizontal black dashed lines show the theoretical limits of the doublet ratio.  \textbf{Bottom right:} \modify{Intrinsic line width (corrected for instrumental resolution) of \civ~absorbers as a function of \ewciv.} The black contour lines indicate the 5th, 25th, 50th, 75th, and 95th percentiles of the distribution.}
    \label{fig:civ_properties}
\end{figure*}

\begin{figure}
    \centering
    \includegraphics[width=0.975\linewidth]{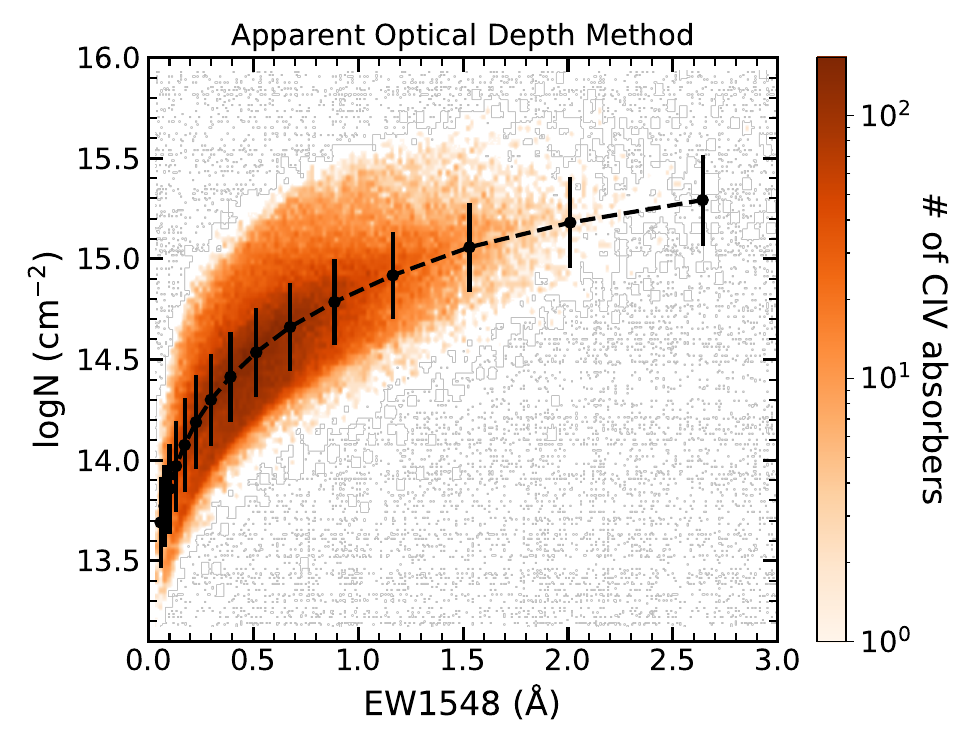}
    \caption{\modify{Total Column density (as measured using the apparent optical depth method) vs \ewciv. The solid circles indicate the mean column density and its standard deviation in redshift bins.}}
    \label{fig:col_dens_civ}
\end{figure}

\begin{figure*}
    \centering
    \includegraphics[width=0.975\linewidth]{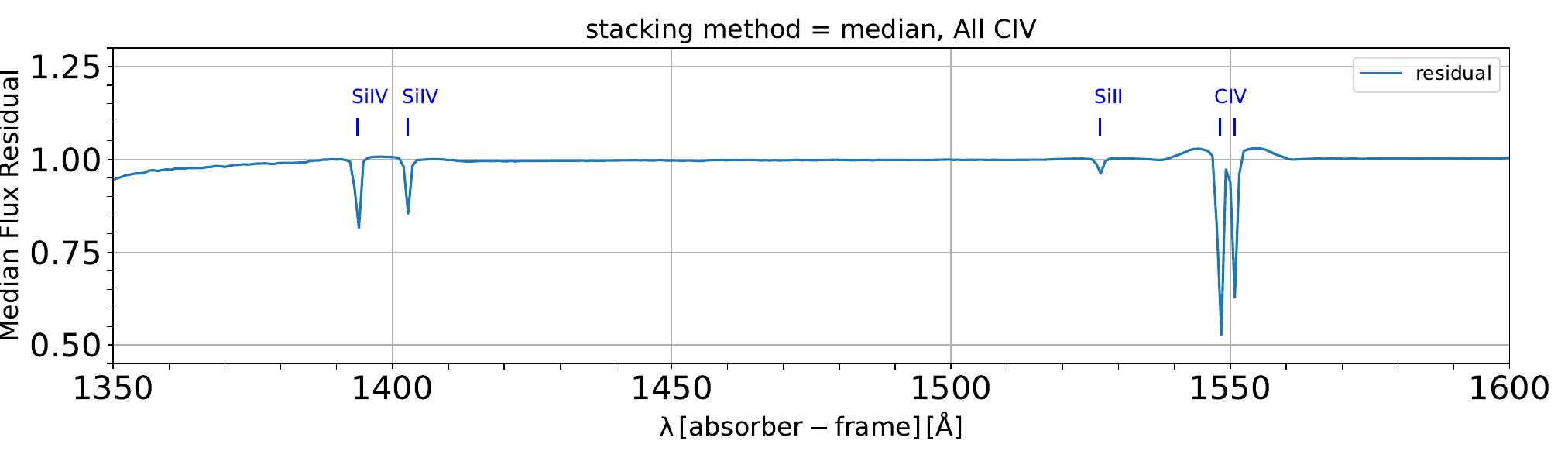}
    \caption{Median composite spectrum of quasars in the rest-frame of detected \civ~absorbers. The other weak metal lines, such as Si~\textsc{ii} and Si~\textsc{iv}, are visible. The spectrum is very flat, indicating good continuum modeling and normalization.}
    \label{fig:civ_stack_spectrum}
\end{figure*}

\section{CIV Absorber Catalogue}\label{civ_catalog}

We ran our absorber detection pipeline on DESI DR1 and DR2 quasars. The pipeline identified \( 32,321 \) and \( 101,487 \) unique \civ~absorber systems from \( 94,986 \) DR1 and \( 300,637 \) DR2 quasars, respectively, a $\sim 33\%$ detection rate, within our wavelength search window. The sample statistics are summarized in Table~\ref{tab:sample_comparison}, along with a comparison to the previous catalog from SDSS. Our sample sizes are $2-7$ times larger than the SDSS DR7 catalog. In the following sections, we present results using only DR2 data to achieve better accuracy on our measurements. Figure~\ref{fig:desispectra} shows an example spectrum where the pipeline detects two \civ~absorbing systems in a high-SNR quasar spectrum. The top panel shows the observed spectrum (blue) and the measured continuum (orange), while the next two panels display the residual spectrum. The continuum appears well-fitted and models the intrinsic quasar features reasonably accurately. The quasar residual is fairly flat around $R\sim 1$, indicating a good continuum normalization. In the zoomed-in version, we also display the best-fit double Gaussian used to measure the properties of these systems.

\subsection{Properties of Individual \civ~Systems}\label{properties}

Figure~\ref{fig:civ_redshift} shows the overall redshift distribution of detected \civ~absorbers in DR1 (orange) and DR2 (blue) quasars. The distribution appears to follow a \modify{right-skewed distribution} between \( 1.3 < z < 4.5 \), with fewer detections near the edges (noise is large near the spectral edges, and also we have very few quasars at high redshifts), making absorber identification more challenging. The distribution peaks around \( z \sim 2.2 \), similar to UV background emission (more discussion in section~\ref{discussion}). 

Next, in Figure~\ref{fig:civ_properties}, we present the properties of detected \civ~systems. The top left panel shows the observed distribution (blue) of the rest-frame equivalent widths (\ewciv) of the \civ~doublet. In the top right panel, we compare the rest-frame equivalent widths of both lines. A positive correlation between the equivalent widths is evident, as expected. Strong absorber systems exhibit higher equivalent widths for both lines, indicating more flux loss in the spectrum. The dashed black lines show the $y=x$ and $y=x/2$ lines derived from the oscillator strength ratio and indicate that the majority of the systems lie within that region. In the bottom left panel, we show the measured doublet ratio (\( DR \)) of \civ~absorbers as a function of \(\mathrm{EW}_{\mathrm{CIV}}\). We find that most systems ($\sim 87.5\%$) fall within the theoretical limits of \( 1 \leq DR \leq 2 \), expected because of the selection criteria. The strong absorbers (\civewGr{1.2}) tend to be saturated\footnote{This follows from the fact that in optically thin regions (\(\tau \ll 1\)), the rest-frame equivalent width (EW) is proportional to the oscillator strength. In optically thick regions (\(\tau \gg 1\)), absorption reaches a maximum for both lines, making it independent of oscillator strength. As a result, both lines exhibit the same EW, leading to \( DR = 1 \) \citep{draine11}.} (\( DR \sim 1 \)). In contrast, weak absorbers ($EW_{\rm 1548}<0.6\,\rm \AA$) are mostly unsaturated (\( DR \sim 2 \)). This also indicates that the majority of our absorbers are genuine systems. \modify{In the bottom right panel, we show the intrinsic velocity dispersion (corrected for instrumental resolution) of \civ~systems as a function of \ewciv. The median intrinsic line width of \civ~absorbers in our sample is approximately $60~\mathrm{km\,s^{-1}}$.} The solid contours represent the 5th, 25th, 50th, 75th, and 95th percentiles, and clearly indicate that most of our data lies within the expected theoretical limits.

\subsection{Column Densities}\label{column_density}

To measure the total column densities ($\mathcal{N}$) of our absorbers, we make use of the apparent optical depth method (AODM; \citealt{savage1991}). This method converts \civ~absorption-line profiles into profiles of apparent optical depth (\( \tau \)) and integrates over a velocity range (in this study, we use $\delta v = \pm 300\,\rm km/s$) to estimate apparent column densities. AODM is known to provide only lower limits for saturated absorbers (\( DR = 1 \); \citealt{cooksey10}) and may systematically overestimate true column densities in low-SNR spectra \citep{fox2005}. However, it remains a reliable approach for estimating column densities in low-resolution spectra and has been widely used in previous studies \citep{cooksey10, cooksey13, mathes2017}.

With \civ~doublets, we have two separate measurements for each system and follow a similar approach as implemented in \citet{cooksey10}. For unsaturated absorbers, i.e., $DR \geq 2 - \sigma_{\rm DR}$, we measure the total column density as the inverse-variance weighted mean of both measurements. \modify{We label the column densities as lower limits for systems classified as partially saturated, defined as those with doublet ratios satisfying $DR < 2 - \sigma_{\rm DR}$. Most absorbers ($\sim88.7\%$) are partially saturated in our sample by this criterion; this is expected as weak systems are very difficult to detect in low-resolution, low-SNR spectra. This is also quantitatively described in section~\ref{completeness}.} 

\modify{In these cases, we adopt the column density measured from the weaker $\lambda1550$ transition, which is less affected by saturation due to its lower oscillator strength. This provides a more stringent and reliable lower limit on $\mathcal{N}$ when unresolved saturation is present. For partially saturated systems where the measured difference $\log \mathcal{N}(1550) - \log \mathcal{N}(1548)$ lies within the range tabulated in Table 4 of \citet{savage1991} (also see, \citealt{jenkins1996}), we apply their empirical correction to recover the true column density.}

\modify{Additionally, we account for a 5\% systematic uncertainty in continuum normalization (see also Figure~\ref{fig:residual_qso_frame}), which is propagated as an additional error in the optical depth calculation, following the prescription in Eqn.~14 of \citet{savage1991}. This systematic is added in quadrature with the statistical uncertainty in the flux when calculating the error on optical depth. The total uncertainty in column density thus includes both statistical and systematic contributions.}

We show the column density as a function of \ewciv in  Figure~\ref{fig:col_dens_civ}. The column density varies from \modify{$13.5\leq \mathrm log[\mathcal{N}/cm^{2}]\leq 15.5$} for our catalog, with the typical errors of \modify{$\sigma_{\mathrm log\mathcal{N}} \sim 0.12$}. \modify{The mean column density and its standard deviation in each redshift bin are shown as black circles.}

\subsection{Properties of \civ~Systems in Composite Spectum}\label{stacking}

Stacking is a powerful tool for detecting weak systems that are otherwise difficult to find due to the low SNR of individual spectra. We perform a pixel-wise median stacking of quasar residual spectra in the rest frame of detected \civ~systems to reveal features that become visible in the stacked spectrum. We chose the median method because it is more stable than the mean to outlier residuals. In Figure~\ref{fig:civ_stack_spectrum}, we present the composite median spectrum of parent quasars stacked on all \civ~absorbers in our catalog. Several weak absorption transitions, such as Si~\textsc{ii} and Si~\textsc{iv}, are evident in the composite spectrum. The presence of multiple transitions in the stacked spectrum indicates that most absorbers detected by our algorithm are real systems. As previously described, since our continuum fitting was limited to a narrow wavelength range outside the \mgii~region, we do not detect \mgii~absorbers in the composite spectrum \citep[see][for a full composite spectrum]{anand2021}. 

One key observation is that our composite spectrum remains nearly flat (albeit, \(\sim 1-3\%\) variation from \( R=1 \) in regions near the \civ~absorption features) almost everywhere, further confirming that \texttt{picca} continua combined with additional median filtering perform well on our quasar sample. However, the continuum underestimation near the \civ~region may introduce a systematic bias of \( \lesssim 3\% \) in the measured equivalent widths. Given that the typical statistical uncertainties on our EWs are \( 5-10\% \) of the total equivalent width, this systematic effect does not dominate our measurements. We also find that absorption features in the stacked spectrum can be well-modeled by single or double Gaussian profiles. \modify{This is because, stacking enables the detection of intrinsically Gaussian absorption features resulting from physical broadening mechanisms. The stacking process itself does not alter the intrinsic shape of the absorption feature.}

The composite spectrum should reflect the average properties of absorbers if they were genuine. Any discrepancy between the composite and individual measurements can indicate the presence of false positives in the catalog. This concept provides a rough estimate of the catalog's purity \citep{pieri2010, pieri2014}. It is important to note that this simple approach could provide an \modify{upper estimate} on the purity and may not account for all potential sources of error or bias in the catalog. To assess the purity of our catalog, we first measure the equivalent widths of \civ~lines in the composite spectrum and compare them with the mean equivalent width in our catalog. The ratio of these two quantities serves as a rough upper estimate of the catalog's purity. Using this approach, we find a purity of \( \sim 95\% \) for our catalog for \civewGr{0.2} systems. However, the actual purity also depends on the strength and redshift of absorbers, similar to completeness \citep[see][for more details]{anand2021}. Additionally, since our detection method incorporates the physical properties and SNR of \civ~doublets, we expect the catalog to be highly pure. As previously shown in Figure~\ref{fig:civ_properties}, the distribution of equivalent \modify{width} measurements and doublet ratio also indicates that the majority of the systems in our catalog are genuine.

\begin{figure*}
    \centering
    % Left Panel (2D Heatmap)
    \begin{minipage}{0.55\textwidth}  % Left panel width
        \centering
        \includegraphics[width=0.99\linewidth]{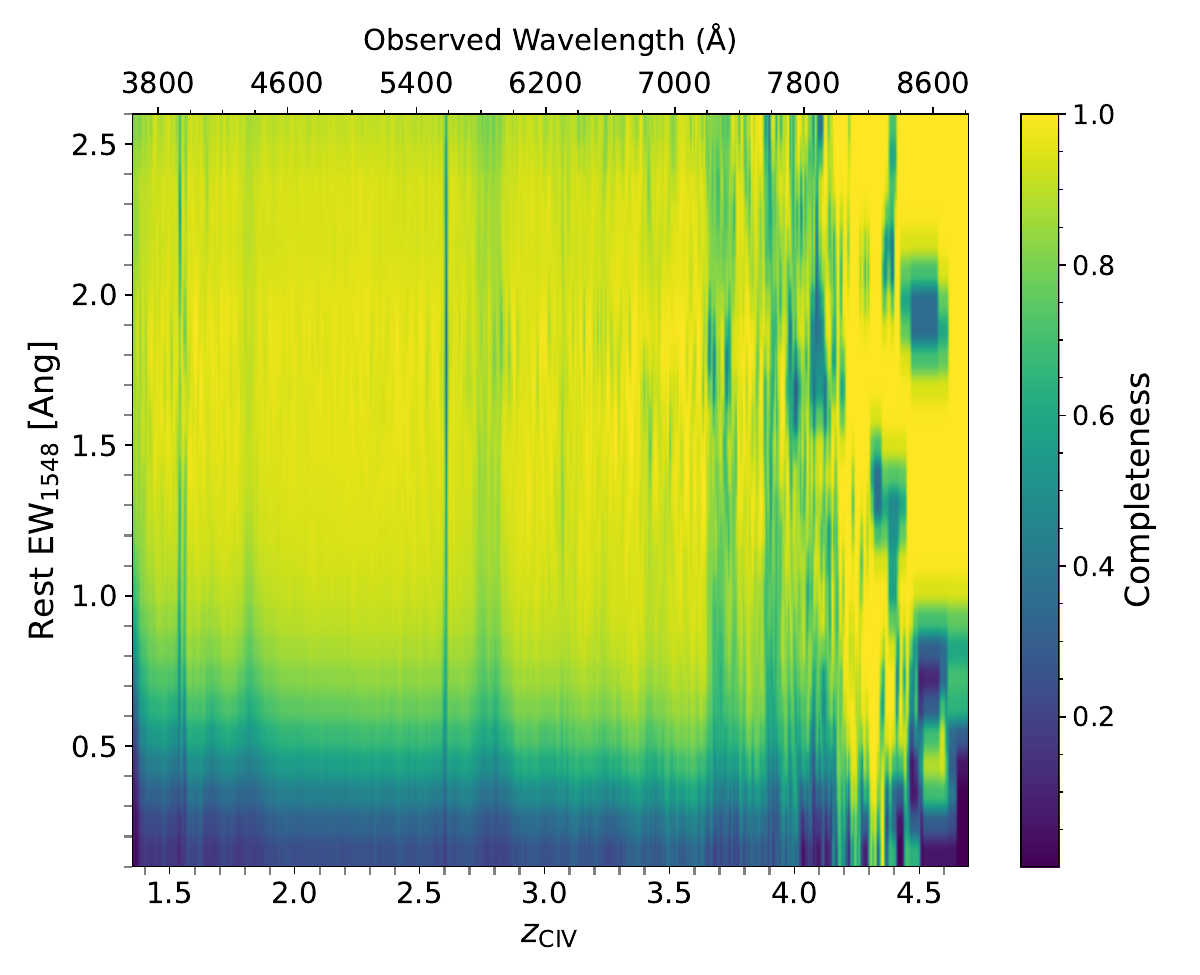}
    \end{minipage}
    % Right Panel (Three stacked plots)
    \begin{minipage}{0.425\textwidth}  % Right panel width
        \centering
        \includegraphics[width=\linewidth]{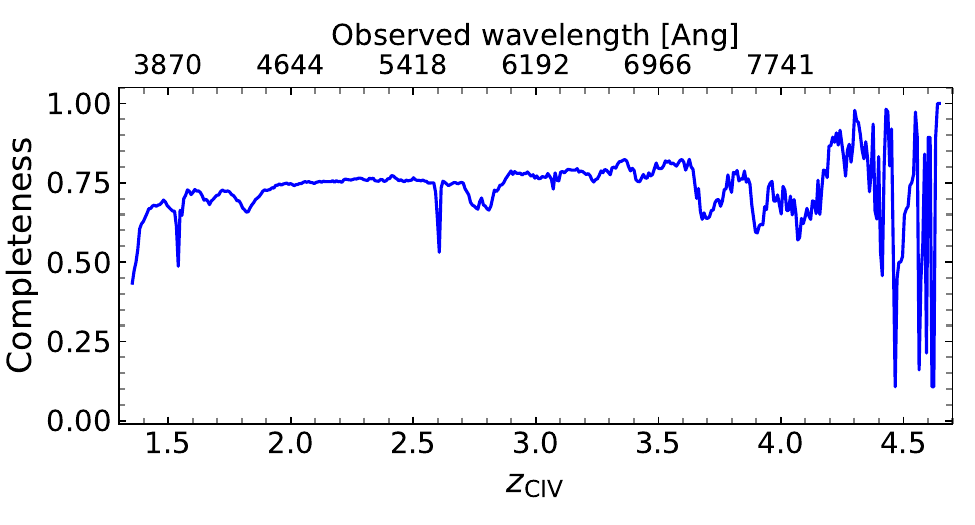}
        \includegraphics[width=\linewidth]{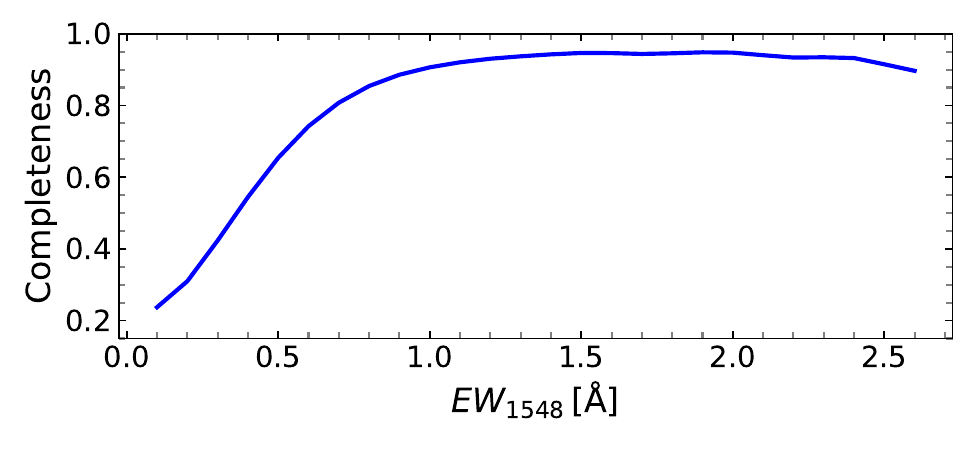}
    \end{minipage}
    \caption{Completeness across different parameters. \textbf{Left:} 2D $C(EW,z)$ completeness heatmap as a function \ewciv and $z$. The low completeness in the masked wavelength and sky line regions are clearly visible. \textbf{Right:} Completeness trends with \civ~redshift and equivalent width.}
    \label{fig:completeness}
\end{figure*}

\subsection{Completeness of CIV Absorbers}\label{completeness}

Detecting absorber systems in noisy data, such as DESI and SDSS, is challenging, making it essential to accurately characterize the efficiency and reliability of any detection algorithm. We follow the Monte Carlo simulation approach described in \citealt{anand2021}. Our goal is to generate double Gaussian profiles that mimic the \civ~absorber profile and insert them into real residual spectra at random pixels. To construct a simulated doublet, we randomly select \(\mathrm{EW}_{\mathrm{CIV}}\) from a uniform distribution within \( 0.1 < EW_{\rm 1548} < 3 \)~\AA, and the doublet ratio (\( DR \)) from a uniform distribution in the range \( DR \in U[1,2] \). Furthermore, we choose the amplitude for the \( \lambda1548 \) line from the range \( A_{\rm 1548} \in U(0, 1) \), ensuring that the rest-frame, instrumental-resolution-corrected line width satisfies \( \sigma_{\rm 1548} \geq 0 \). We adopt the same line width (\( \sigma \)) for both components of the doublet. Using these sampled parameters, we generate synthetic doublet Gaussian profiles and insert them into the real residual spectra. Additionally, the associated error arrays from the original residual spectra are used during this process. While inserting these absorbers, we ensure proper wavelength masking so that the simulated absorbers are consistent with the real detection pipeline wavelength range.  

We then apply our detection method, including SNR cuts (\civ~\textit{doublet SNR criteria}), to the modified residual spectra. A mock absorber is considered detected if it satisfies the \civ~\textit{doublet SNR criteria}, the velocity difference between the measured and inserted redshift satisfies\footnote{This corresponds to 3–8 times the DESI spectral resolution and allows for a reasonable measurement of absorber redshifts.} \( |\Delta v| \leq 500 \,\mathrm{km\,s^{-1}} \), and the fractional change in equivalent width meets the condition \( |EW_{\rm inserted} - EW_{\rm measured}| < EW_{\rm inserted} \). In total, we simulated approximately \( \sim 3 \) million mock absorbers in \( \sim 300,637 \) residual spectra. This large sample size ensures that uncertainties in the completeness analysis are negligible and that the completeness function exhibits a smooth trend as a function of redshift and \ewciv. 

Finally, we estimate the two-dimensional completeness function,  
\(C(EW, z) = \frac{n_{\rm det}}{n_{\rm ins}}\)
in fine bins of \ewciv and \( z \), where \( n_{\rm det} \) is the number of \textit{detected} \civ~absorbers and \( n_{\rm ins} \) is the number of \textit{inserted} absorbers. Using this \( C(EW,z) \), we calculate the completeness-corrected number of $i^{\rm th}$ absorber, falling in $(EW, z)$ bin, as  
\(N_{\rm corr, i} = \frac{1}{C_{i}(EW, z)}\). This corrected number represents an estimate of the absorber counts that would have been detected if our detection method were perfect. The left panel of Figure~\ref{fig:completeness} shows the two-dimensional completeness function \( C(EW, z) \). We find that our measured completeness is a strong function of both \(\mathrm{EW}_{\mathrm{CIV}}\) and \( z \). Strong absorbers are easier to detect, except near the higher end of redshift, where they are diluted by noise. In contrast, weak absorbers remain difficult to identify across the entire spectrum. We also observe low completeness in regions masked by our search method. The lower completeness near the masked calcium lines ($\lambda~\sim 4000$~\AA) and other skylines ($\lambda \sim 5600,\, 6300~$\AA) are clearly visible. Since there are very few quasars in our sample at \( z > 4.1 \), the number of mock absorbers in this redshift range is also relatively low, matching the distribution of real absorbers. As a result, our completeness estimates are noisier in this region.

In the right two panels of Fig~\ref{fig:completeness}, we present the averaged completeness as a function of \civ~absorber properties. The upper panel shows the completeness as a function of \(\mathrm{EW}_{\mathrm{CIV}}\), averaged over all redshifts. The increasing completeness with \(\mathrm{EW}_{\mathrm{CIV}}\) is clearly visible, as strong absorbers are easier to detect. Our method is approximately \( 50\%\) complete at \( EW_{\rm 1548} \sim 0.4 \)~\AA\, and reaches \( \gtrsim 95\%\) completeness for \( EW_{\rm 1548} \gtrsim 1.5 \)~\AA. In the bottom panel, we present completeness as a function of \civ~redshift, averaged over all \(\mathrm{EW}_{\mathrm{CIV}}\). The dips at specific redshifts correspond to wavelength regions masked during the search. Finally, in Figure~\ref{fig:qso_completeness} (see, Appendix~\ref{qso_residual}), we examine the dependence of completeness on the quasar rest-frame wavelength. We observe that completeness remains relatively stable at \( \sim 70-80\%\) across most wavelengths, except at the edges. This indicates that our continuum estimates are fairly accurate, as the variation remains nearly flat, consistent with expectations for robust continuum modeling.

Finally, we present the observed and completeness-corrected \(\mathrm{EW}_{\mathrm{CIV}}\) distributions in Figure~\ref{fig:civ_properties} (top left) for all \civ~absorbers detected with our algorithm. The completeness-corrected distribution (orange) appears to follow a smooth exponential profile, suggesting that the apparent turnover at \( EW_{\rm 1548} \sim 0.5 \)~\AA\ in the observed distribution (blue) is not real but rather a consequence of missing many weak systems in our search. This result is consistent with the \mgii~and \civ~absorber distributions observed in several previous studies \citep{nestor05, cooksey10, zhu13a, anand2021}. This is expected, as weak systems are naturally lost in low-resolution, low-SNR spectra due to signal loss in noisy regions. In the results section, we apply completeness correction to perform all subsequent analyses.

\subsection{Comoving Redshift Path of the Survey}\label{surveypath}

The absorber sensitivity function, \( g(EW_{\rm r}, z) \), estimates the total number of quasar spectra in the survey where a \civ~absorber with rest-frame \( EW_{\rm r} \) or greater could be detected at redshift \( z \). We follow the mathematical definition of \( g(EW, z) \) as proposed in the literature \citep{nestor05, cooksey10, hasan20}, where this function is computed as a sum over all \( N_q \) quasars in the sample as follows:
\begin{equation}
    g(EW_{r}, z) = \sum_{i=1}^{N_q}H(z-z_{i, min})H(z_{i, max}-z)
\end{equation}
where \( H(x) \) is the Heaviside function, and \( z_{i, \text{min}} \) and \( z_{i, \text{max}} \) denote the minimum and maximum \civ~redshifts that could be detected in the \( i \)th quasar residual and $N_q$ is the total number of quasars in the sample. These limits are determined based on the search window defined in the previous section. In this formulation, we assume absorbers with any strength (a very conservative condition) can be detected in a given residual at unmasked pixels within our method. 

Using \( g(EW_{\rm r}, z) \), we can now define the comoving path density\footnote{In the literature, this is sometimes referred to as absorption distance.}, \( \Delta X \), which quantifies the total path length covered by the survey within a finite redshift interval \( z_{1} \) and \( z_{2} \). Mathematically, it is defined as:
\begin{equation}
    \label{eqn:pathdens}
    \Delta X = \int_{z_{1}}^{z_{2}}g(EW_r, z)\frac{dX}{dz}dz
\end{equation}
where, assuming flat Universe, $dX/dz$ is given by \citep{bahcall1969}\footnote{The probability of finding an absorber (with number density $n(z)$ and cross-section $\sigma$) along a line of sight in a cylindrical volume is given by $dP = \pi R^{2}n(z)(1+z)^2\frac{cdz}{H(z)} = \frac{c}{H_{\rm 0}}n(z)\sigma\frac{dX}{dz}dz$; where $H(z) = \sqrt{\Omega_M(1+z)^3+\Omega_{\Lambda}}$, assuming a flat Universe.}
\begin{equation}
\begin{split}
\frac{dX}{dz} &= \frac{(1+z)^2}{\sqrt{\Omega_M (1+z)^3 + \Omega_\Lambda}}, \\
\implies X(z) &= \frac{2}{3\Omega_M}\sqrt{\Omega_M (1+z)^3 + \Omega_\Lambda} 
\end{split}
\label{eqn:dXdz}
\end{equation}
with \( \Omega_M \) and \( \Omega_{\Lambda} \) representing the matter and dark energy densities, respectively. More details can be found in \citet{bahcall1969}. 
\begin{figure*}
    \centering
    \includegraphics[width=0.98\linewidth]{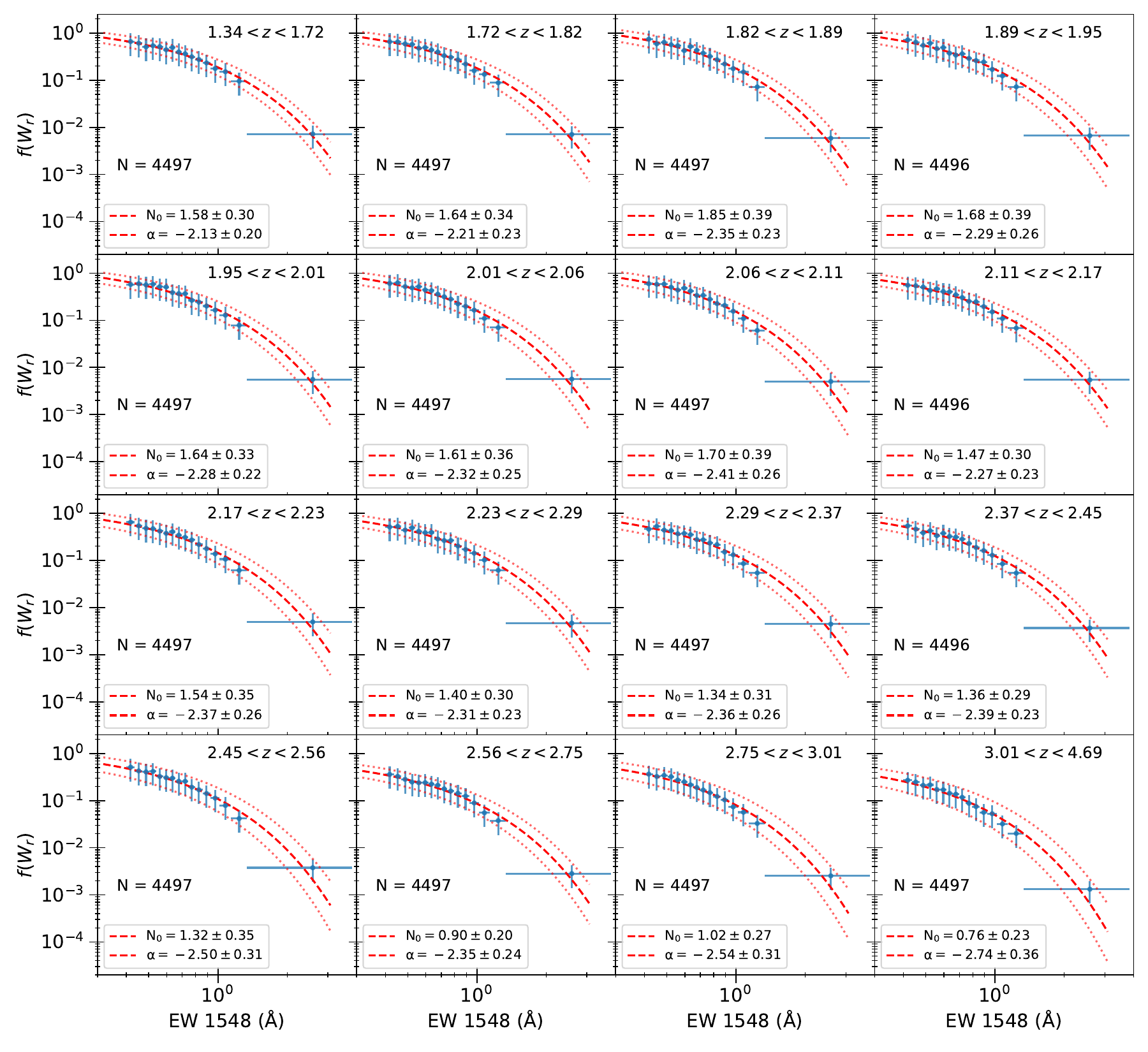}
    \caption{The differential \civ~equivalent width frequency distribution (Eqn~\ref{eqn:fwr}, also called intrinsic incidence rates) is shown as a function of redshift and \(\mathrm{EW}_{\mathrm{CIV}}\). The redshift bins are chosen such that each bin contains at least \( \sim 4500 \) absorber systems. The maximum likelihood best-fit exponential models (Eqn~\ref{eqn:ml_fit}) are shown as dashed lines, with \( \pm 1\sigma \) uncertainties represented by dotted lines. The observations have been corrected for their completeness. \modify{The frequency distribution shows little evolution with redshift.}}
    \label{fig:incidence_rate}
\end{figure*}

\section{Results}\label{results}

The key results that we present in this paper include the differential \civ~equivalent width frequency distribution, comoving path density, and cosmic mass density of \civ~absorbers in the Universe within the redshift range \( 1.4 < z < 4.5 \) for absorbers with \( \mathrm{EW}_{\mathrm{CIV}} \geq 0.4 \)~\AA, using the latest DESI quasar data. We chose this threshold because we are $\gtrsim 50\%$ complete above $\sim 0.4\,\rm \AA$. Moreover, we use quasars with \( z_{\rm qso} \geq 1.7 \) ($\sim 96\%$ of the sample) for the \( \Delta X \) calculation, as used in \citet{cooksey13}. Consequently, we also remove absorbers ($\lesssim 1.5\%$) associated with quasars at \( z_{\rm qso} < 1.7 \) for consistency. Finally, we compare our results with existing literature. This work is the largest statistical analysis of \civ~absorbers to date based on one of the most extensive spectroscopic datasets in astronomy.

\subsection{CIV Equivalent Width Incidence Rates}\label{incidence_rates}

The rest-frame equivalent width incidence rate or the frequency is defined as the true number of absorbers (i.e., completeness corrected) per unit comoving redshift path (estimated in a redshift bin $\Delta z$ using Eqn~\ref{eqn:pathdens}) per unit equivalent width ($\Delta W_r$). Mathematically, it is described by the probability density function \( f(W_{r}) \):
\begin{equation}
\begin{aligned}
    f(W_r) &= \frac{d^2N}{dW_r dX} = \frac{N_{\rm corr}}{\Delta W_r \Delta X} \\
    N_{\rm corr} &= \sum_{i=1}^{N_{\rm abs}} w_i, \quad w_i = \frac{1}{C_i}\\
    \sigma_{f(W_r)} &= f(W_r) \sqrt{\left(\frac{\sigma_{N_{\rm corr}}}{N_{\rm corr}}\right)^2 + \left(\frac{\sigma_{\Delta W_r}}{\Delta W_r}\right)^2}, \\
    \sigma_{N_{\rm corr}} &= \sqrt{\sum_{i} w_{i}^2}, \quad \sigma_{W_r} = \frac{\Delta W_r}{2}
\end{aligned}
\label{eqn:fwr}
\end{equation}
where the second equality (in the first equation) follows from the fact that, in practice, we compute this probability density in discrete bins of \( W_r \) and \( z \), given the finite sample size \citep[see also][]{storrie1996}. The term \( C_j \leq 1 \) represents the completeness factor for the \( j^{\text{th}} \) absorber, \( \Delta W_r = W_{r, i+1} - W_{r, i} \) denotes the bin size of equivalent widths, and \( \Delta X \) is given by Eqn~\ref{eqn:pathdens} and $N_{\rm abs}$ is number of absorbers in that fine 2D equivalent width and reshift bin. The uncertainties on \( N_{\rm corr} \) are estimated using a Poisson distribution approximation \citep{cooksey10,hasan20}.

We present the differential \civ~equivalent width frequency distribution in Figure~\ref{fig:incidence_rate} as a function of rest-frame equivalent width and redshifts. The redshift bins are chosen such that \modify{there are about the same number of systems \( \sim 4500 \) in each of the bins. This ensures that statistical uncertainties remain comparable across bins and helps minimize measurement fluctuations due to varying sample sizes. Additionally, our goal was to strike a balance between having sufficient statistics and avoiding overly fine redshift binning, which could introduce fluctuation.} The vertical and horizontal error bars represent the uncertainties on \( f(W_r) \) and \(W_r \), respectively, as estimated from Eqn~\ref{eqn:fwr}. We find that the equivalent width incidence rate exhibits a steep decline with increasing absorber strength in all redshift bins, indicating that weak \civ~systems are more common in the Universe than their strong counterparts. This trend is consistent with previous \civ~studies \citep{cooksey10, cooksey13, hasan20}. We fit the incidence rate using a maximum likelihood method with an exponential profile (defined below), following the approach used in previous studies \citep{nestor05, cooksey13}:
\begin{equation}
    f(W_r) = N_{\rm o}e^{\alpha W_r}
    \label{eqn:ml_fit}
\end{equation}
where the fitting parameters \( N_{\rm o} \) and \( \alpha \) are simultaneously fitted to obtain the best-fit values. These values and their associated uncertainties are shown in each panel of Figure~\ref{fig:incidence_rate}. The best-fit amplitudes (\( N_{\rm o} \)) and slopes (\( \alpha \)) do not evolve much with redshift, indicating that the differential incidence rate of \civ~absorbers follows a global smooth trend across cosmic epochs. The slopes vary between $-2.13$ to $-2.74$, while the amplitudes change between $0.77$ and $1.71$. We also fitted the same exponential model to the incidence rate for the entire sample and recovered the similar best-fitted parameters ($\alpha = -2.35 \pm 0.25$ and $N_{\rm o} = 1.32 \pm 0.29$). 

Comparing our fitting results with \citet{cooksey10}, we find that our best-fit normalization amplitudes and slopes\footnote{The slopes lie between $-2.49$ to $-2.82$ and amplitudes are well within $1.82$ and $5.1$, though the error bars are quite large in \citet{cooksey13}.} are slightly smaller but consistent within $2\sigma$ uncertainty. This difference is not significant and could be attributed to variations in sample selection, mean signal-to-noise of absorbers, median redshifts, and completeness corrections compared to their catalog. However, our results remain qualitatively consistent with their results.

\begin{figure}
    \centering
    \includegraphics[width=0.98\linewidth]{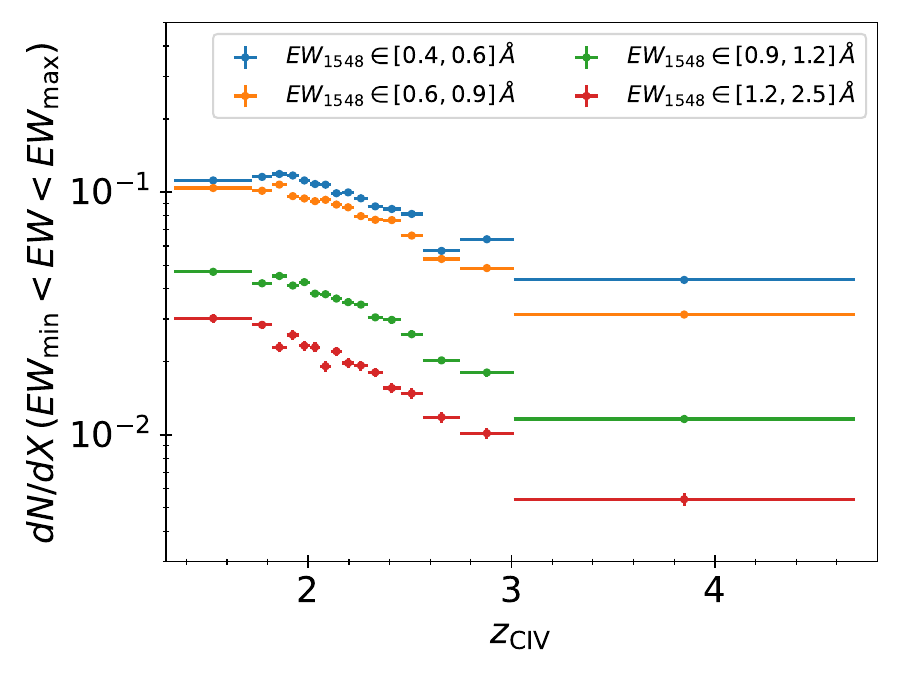}\\
    \includegraphics[width=0.98\linewidth]{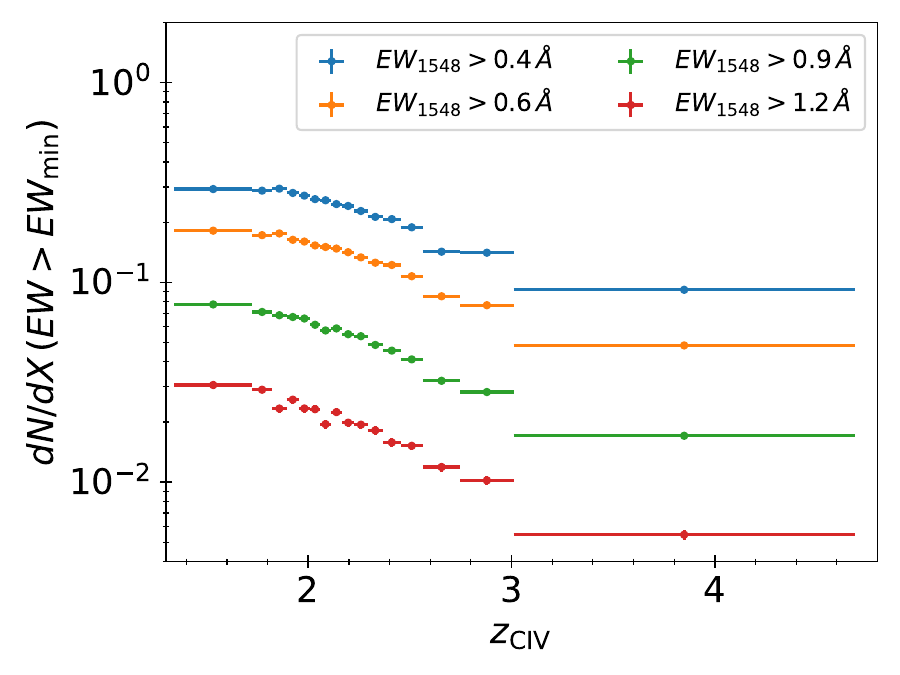}
    \caption{Differential (top) and cumulative (bottom) comoving path density of \civ~absorbers as a function of redshift and \ewciv. The number of \civ~absorbers per unit comoving path length increases smoothly from $z\sim 4.5 \rightarrow1.4$ by a factor of $\sim 2-3$ for all \ewciv bins. Strong absorbers have lower densities than weak systems.}
    \label{fig:dn_dx_rate}
\end{figure}

\subsection{Comoving Path Densities}\label{comovingpath}

Next, we examine the properties of absorbers in terms of comoving path densities\footnote{In the literature, this is sometimes referred to as "absorber line density" or "path density."}, i.e., the number of true absorbers (i.e., completeness corrected) per unit comoving path (\( dN/dX \)) that satisfy a given \(\mathrm{EW}_{\mathrm{CIV}}\) threshold. Mathematically, this quantity is defined as \citep{cooksey10, hasan20}:
\begin{equation}
    \begin{aligned}
        \frac{dN}{dX} (x) &= \frac{1}{\Delta X(z)}\sum_{i}{w_{i}}\, ;\\
        \sigma^{2}_{dN/dX} &= \frac{1}{\Delta X^{2}(z)}\sum_{i}{w^{2}_{i}}\\ 
        \text{where, } x&=EW_{\rm 1548}\geq W_{\rm min} \\
        \text{or, } x&= W_{\rm min}\leq EW_{\rm 1548}\leq W_{\rm max}\
    \end{aligned}
\end{equation}
where \( w_i=1/C_i \) is the completeness correction for each absorber, and the uncertainties on \( dN/dX \) are estimated assuming a Poisson distribution and $\Delta X$ is defined in Eqn~\ref{eqn:pathdens}.  

In Figure~\ref{fig:dn_dx_rate}, we present the results for differential (top panel) and cumulative (bottom panel) \( dN/dX \) for different \(\mathrm{EW}_{\mathrm{CIV}}\) thresholds as a function of absorber redshift from \( z \approx 1.4 \) to \( z \approx 4.5 \). The comoving path density of absorbers shows a smooth increasing trend with decreasing redshift. We find that \( dN/dX \) is at least \( \sim 2-3 \) times smaller at \( z \approx 4 \) than at \( z \approx 1.5 \) for all \(\mathrm{EW}_{\mathrm{CIV}}\) thresholds. For \(\mathrm{EW}_{\mathrm{CIV}} \geq 0.6\)~\AA, cumulative bin, we find \( dN/dX \approx 0.18 \pm 0.003 \) at \( z \approx 1.5 \), compared to \( dN/dX \approx 0.05 \pm 0.001 \) at \( z \approx 4 \). A significant decreasing trend is also observed with increasing \(\mathrm{EW}_{\mathrm{CIV}}\), consistent with the absorber incidence rates shown in Figure~\ref{fig:incidence_rate}.  
\begin{figure*}
    \centering
    \includegraphics[width=0.75\linewidth]{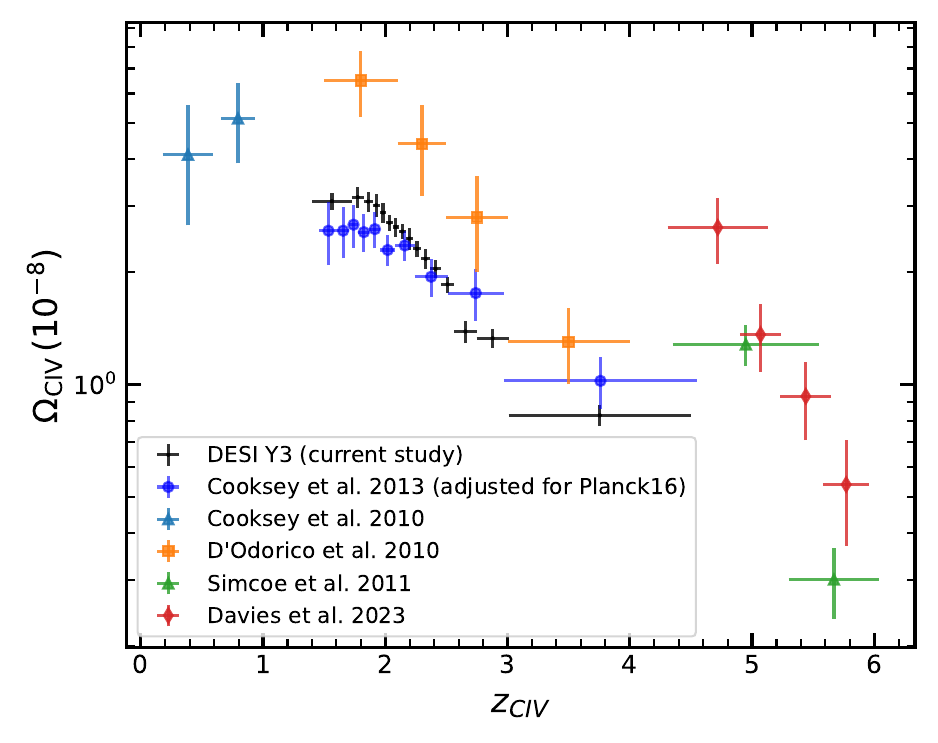}
    \caption{Cosmic mass density of \civ~absorbers as a function of redshift for systems with $\mathrm log\mathcal{N}\geq 14$. Densities measured from SDSS DR7 \citep{cooksey13} (blue, \modify{adjusted for \citealt{planck16} cosmology}) and our catalog (black) are based on column densities estimated using the apparent optical depth method. Since most systems are saturated, these measurements represent lower limits. We complement our latest measurements with results from previous studies \citep{cooksey10, dodorico10, simcoe11, davies2021}. Due to our large sample size, our measurement uncertainties are \modify{very} small. The \civ~cosmic mass density increases smoothly by a factor of \( \sim 2-4 \) from \( z \approx 4 \rightarrow 1.4 \), indicating an overall increase in \civ~abundance as the Universe evolved. We do not adjust previous results to match the \citealt{planck16} cosmology, except for those from \citealt{cooksey13}.}
    \label{fig:mass_density}
\end{figure*}

For the same redshift bin at \( z \approx 4 \), we find \( dN/dX \approx 0.09 \pm 0.001 \) for \(\mathrm{EW}_{\mathrm{CIV}} \geq 0.4\)~\AA, \( dN/dX \approx 0.017 \pm 0.0005 \) for \(\mathrm{EW}_{\mathrm{CIV}} \geq 0.9\)~\AA, whereas for \(\mathrm{EW}_{\mathrm{CIV}} \geq 1.2\)~\AA, the value drops sharply to \( dN/dX \approx 0.005 \pm 0.0003 \), a decline by nearly two order of magnitudes. These trends are in good agreement with those observed in the SDSS-DR7 \civ~absorber sample from \citet{cooksey13}, \modify{highlighting the robustness of our methodology.}

\subsection{CIV Mass Densities}\label{mass_density}

The cosmic mass density or fraction of an element or ion, \( \Omega_{\rm ion} \), is defined as the ratio of the absorber mass per unit comoving path to the critical density of the Universe (\( \rho_{\rm crit} \))\footnote{\( \rho_{\rm crit} = \frac{3H_{\rm 0}^2}{8\pi G} = 8.59\times 10^{-30} \, \rm gm\,cm^{-3} \), where \( H_{\rm 0} = 67.66\, \rm km/s/Mpc \) \citep{planck16}.}. For our sample of intermediate-strength \civ~absorbers (\civewGr{0.4}), we define the \civ~mass density, \( \Omega_{\rm CIV} \)\footnote{As described earlier, \( dN/dX \propto n(z) \sigma(z) \), from which the mass density ($\rho$) can be obtained using column densities ($\mathcal{N}$) as: \(
\rho \propto \mathcal{N} \frac{dN}{dX}\).}, as:
\begin{equation}
    \begin{aligned}
        \Omega_{\rm CIV} (z) &\approx \frac{H_{\rm 0}m_{\rm CIV}}{c\rho_{\rm crit}}\frac{\mathcal{N}_{\rm tot} (z)}{\Delta X} \\
        \text{where, } \mathcal{N}_{\rm tot}(z) &=  \sum_{i=1}^{N_{\rm abs}}w_{i}\mathcal{N}_{i}\,,\,\, \, w_{i} = 1/C_i \\
        \text{and, } \sigma_{\Omega_{\rm CIV}}&=  \Omega_{\rm CIV}\sqrt{\frac{\sum_i w_i^2\sigma_{\mathcal{N}_i}^2}{\mathcal{N}_{\rm tot}^2} +\frac{\sum_{i}w_i^2}{(\sum_iw_i)^2}}
    \end{aligned}
    \label{eqn:mass_density}
\end{equation}

where \( H_{\rm 0} \) is the present-day Hubble constant, \( m_{\rm CIV} = 1.99 \times 10^{-23} \,\rm g \) is the mass of a carbon atom, \(\rho_c = 8.59 \times 10^{-30} \,\rm g\,cm^{-3}\) is the present-day critical density of the Universe, assuming \citet{planck16} cosmology and \( c \) is the speed of light in \(\rm km\,s^{-1} \). The terms \( \mathcal{N}_{i} \) and \( w_{i} \) represent the column density and completeness factor for the \( i^{\text{th}} \) absorber, respectively and $N_{\rm abs}$ total number of absorbers in $\Delta z$ bin. The first and second error terms are related to uncertainty in column densities and number counts, respectively. The errors on number counts are measured assuming Poisson statistics. Note that the total error budget may still be higher than our estimates because the AODM method is known to underestimate column density errors \citep{dodorico10}. 

We present our measurements of \( \Omega_{\rm CIV} \) as a function of redshift for systems with \( \mathcal{N}_{\rm CIV} \geq 10^{14}\, \rm cm^{-2} \) in Figure~\ref{fig:mass_density}. The \civ~cosmic mass density shows a strong declining trend with redshift, suggesting that the amount of carbon in the Universe has increased as it has evolved. At \( z \sim 4 \), we find \modify{\( \Omega_{\rm CIV} = (0.82 \pm 0.05)\times 10^{-8} \)}, while at \( z \sim 1.5 \), it increases to \modify{\( \Omega_{\rm CIV} = (3.16 \pm 0.2)\times 10^{-8} \)}, representing more than threefold \modify{(\( \sim 3.8 \))} increase over \( \sim 3 \) Gyrs. We also compare our results with previous studies across different epochs, which are consistent with the searched redshift range. Our large sample size allows us to tightly constrain (lower limits) the \civ~mass density with such high signal-to-noise. These measurements are complementary to previous studies that have focused on lower or higher redshifts using high or medium-resolution small surveys of \civ~absorbers. Also, given the increased statistics (see Table~\ref{tab:sample_comparison}), our measurement errors are extremely small, and we push to slightly higher redshifts with finer bins. 

It is important to note that our DESI measurements provide \textit{lower} limits, as many of our absorber systems are saturated. A direct comparison should be made with SDSS-DR7-based results from \citealt{cooksey13} (adjusted for Planck cosmology), as they were obtained using a similar instrumental resolution and column density estimation method. On the other hand, the quasar selection functions differ significantly \citep[see][for more details]{chaussidon2023} between SDSS and DESI, and there are many more faint quasars observed with DESI.

We find that our DESI measurements are in qualitative agreement with previous results, though our values are slightly higher than those from SDSS DR7. Several factors could contribute to this discrepancy: 1) The quasar selection functions differ significantly. In DESI, most quasars lie at \( z > 2.1 \), as they have been targeted to detect Ly$\alpha$ forest and have systematically lower spectral signal-to-noise than SDSS DR7 quasars used for \civ~detection in \citet{cooksey13}. 2) The completeness factors vary due to differences in wavelength coverage, mean SNR, and redshift distributions, all of which influence these measurements. The combination of these effects impacts the detection rate in our method ($\sim 33\%$), which is lower than \citet{cooksey13}, which was around $\sim 50\%$. \modify{3) We also apply the column density corrections prescribed in \citet{savage1991} for partially saturated absorbers, which bring our values closer to the true column densities.}

Finally, we also re-measured \( \Omega_{\rm CIV} \) from the \civ~catalog of \citet{cooksey13} using Planck cosmology. The revised values are systematically higher by \( 10-20\%\) (see Figure~\ref{fig:cooksey_omega}, red points) compared to those derived using WMAP cosmology. The reason is that the baryon fraction is higher and the Hubble constant is lower in Planck cosmology, which reduces the absorber path length, $\Delta X$, therefore increasing the mass densities. Overall, our trends are in good qualitative agreement with the SDSS-based results. However, a detailed comparison of these differences and their impact on some of these measured quantities would be interesting in the future. 

Next, we also measure the metallicities using our \( \Omega_{\rm CIV} \) estimates using carbon as a metallicity tracer in the IGM. As pointed out in \citet{ryanweber2009} and adapted in \citet{manuwal2021}, \( \Omega_{\rm CIV} \) can be used to estimate lower limits on IGM metallicities (\( Z_{\rm IGM} \)) traced by carbon using the relation
\begin{equation}
Z_{\rm IGM} = \frac{\Omega_{\rm CIV}}{\Omega_{\rm b} A_{\rm C} f_{\rm CIV}} 
\end{equation}
where \( A_{\rm C} \) is the mass fraction of carbon in total metals, \( f_{\rm CIV} \) is the fraction of \civ~to total carbon, and \( \Omega_{\rm b} = 0.049 \) is the baryon density in the Universe \citep{planck16}. We adopt a constant value of \( A_{\rm C} = 0.178 \) over our redshift range, following \citet{ryanweber2009}, and assume \( f_{\rm CIV} \leq 0.35 \) \citep[see Fig. 10 of][]{oppenheimer06}, corresponding the peak fraction of \civ~absorbers. For solar metallicity, we adopt a value of $Z_{\rm \odot}=0.0122$ \citep{asplund2005}. 

Since our cosmic mass density measurements provide only lower limits, we derive conservative lower limits on the IGM metallicity, finding \( \log (Z_{\rm IGM} / Z_{\rm \odot}) \gtrsim -3.25 \) at \( z \sim 2.3 \) and \( \log (Z_{\rm IGM} / Z_{\rm \odot}) \gtrsim -3.7 \) at \( z \gtrsim 4 \), which are consistent with previous studies \citep{schaye2003, ryanweber2009, simcoe11,madau14,ahvazi2024}, and further indicates that overall the mean IGM metallicity has increased by a factor $\sim 3$ in this redshift range.

%%%%%%%%%%%%%%%%% DISCUSSION %%%%%%%%%%%%%%%%%%

\section{Discussion}\label{discussion}

The analysis of hydrogen and metal absorbers detected in quasar spectra provides crucial insights into the baryon cycle, the cosmological evolution of matter and its connection to large-scale structure, and the ionization history of the Universe across a broad range of redshifts \citep{reddy2009, madau14, bouwens2015}.  

In general, one of the key goals of the field is to compare the cosmic evolution of different metal species and their ionization conditions to better understand the chemical evolution of the Universe. Neutral hydrogen is among the most extensively studied absorbers, with its cosmic mass density evolution characterized over a wide range of redshifts \citep{songaila05, ho2021}. Similarly, the cosmic mass density of metal absorbers such as \mgii~\citep{mathes2017, codoreanu2017, davies2021, abbas2024}, \civ~\citep{peroux04, songaila05, cooksey13}, and Si\,\textsc{iv} \citep{songaila01, cooksey2011, shull2014} has been explored in previous studies. 

In this paper, we present the evolution of the \civ~mass density (\( \Omega_{\rm CIV} \)) and IGM metallicities (traced by carbon) using the largest dataset to date, obtained from an ongoing Stage-IV spectroscopic mission. The large sample size enables us to trace \civ~path densities, incidence rates, and mass densities over a broader redshift and equivalent width range with higher SNR measurements and improved cosmological parameters.

%start here
\subsection{Cosmic Evolution of CIV absorbers at $1.4<z<4.5$}\label{evolution}

Since the first survey of \civ~absorbers detected in background quasars \citep{sargent88}, several studies have revealed that the absorber path density, \( dN/dz \), increases by a factor of \( \sim 3-4 \) from \( z \sim 4.5 \) to \( z \sim 1.4 \) \citep[see also][]{steidel1990, misawa2002, peroux04} for intermediate strength absorbers (\(\mathrm{EW}_{\mathrm{CIV}} \geq 0.3\)~\AA). Recent studies have quantified the incidence of \civ~absorbers in terms of comoving path density (\( dN/dX \)), which represents the true number of systems expected along the absorber path covered in a given survey. It is also directly proportional to the product of the underlying cosmic number density, $n(z)$, and absorber cross-section, $\sigma(z)$, at any given epoch. 

For a non-evolving absorber population, this product remains constant\footnote{Although the universe expands, the comoving volume remains constant because comoving distances account for cosmic expansion. This is also related to the probability of finding an absorber in a given infinitesimal comoving volume. }; therefore, \( dN/dX \) should not change with redshift. However, the redshift path density \( dN/dz \) could still change \citep{nestor05, hasan20, hasan21}. Therefore, if $dN/dX$ evolves with redshift, this implies that either of them or both must change over cosmic time.

As shown in Figure~\ref{fig:dn_dx_rate}, we find that $dN/dX$ has increased by a factor of $2-5$ from \( z=4.5 \rightarrow 1.5 \), across all absorber strengths from weak ($EW_{\rm 1548}<0.6\,\rm \AA$) to strong (\civewGr{1.2}, see differential values in the top panel and cumulative in the bottom panel). These trends are in good agreement with previous studies and suggest that either the absorber physical cross-section or the cosmic number density ($n(z)$), or both, or their combination, have increased over this redshift range. \citet{hasan21} estimated the radius of \civ-absorbing clouds by assuming that the \civ~cosmic number density is the same as that of galaxies with \( L \geq 0.01L_{\star} \). Their findings indicate that the gas radius increases by a factor of \( \sim 2 \) from \( z \approx 4 \) to \( z \approx 1.5 \) for absorbers \civewGr{0.3}. Consequently, the absorber cross-section increases by a factor of \( \sim 4 \), leading to a corresponding increase in \( dN/dX \) by nearly the same amount. This result aligns quite well with our observational analysis.

Moreover, \citet{peroux04} found that the \civ~doublet ratio declines as the Universe grew, indicating an overall increase in the mean \civ~column density ($\braket{\mathcal{N}}$) and carbon abundance from $z=4.5\rightarrow 1.4$, also contributing to the increasing trend of $\Omega_{\rm CIV}$, as cosmic density is also directly proportional to $\braket{\mathcal{N}}$ (see Eqn~\ref{eqn:mass_density}). This evolution is likely also connected to the overall star formation history, which has increased by a factor of \( \sim 3 \) from \( z \approx 4 \rightarrow 2 \) \citep{madau14}. As stars evolve, they synthesize carbon, which is then expelled into the IGM through powerful winds or supernova-driven outflows \citep{rauch2001, becker2009}.

It is also important to recognize that \( \Omega_{\rm CIV} \) and \( dN/dX \) are complementary measurements, each influenced by different absorber populations. Since \( \Omega_{\rm CIV} \) is derived from total column densities, it is primarily dominated by strongly saturated absorbers in low-resolution surveys. In contrast, \( dN/dX \) is sensitive to weak absorbers ($EW_{\rm 1548}<0.6\,\rm \AA$), which are far more prevalent, as indicated by their low completeness factor (see Figure~\ref{fig:completeness}). As described above, since \( \sigma(z) \) increases by a factor of \( \sim 4 \) in this redshift range, and the total column density can also be estimated from the first moment of the column density frequency function, \( f(\mathcal{N}) = \frac{d^{2}N}{d\mathcal{N}dX} \) \citep{abbas2024}, it follows that \( \Omega_{\rm CIV} \) should exhibit a similar trend. This is consistent with our observations in Figure~\ref{fig:mass_density}, where the cosmic mass density smoothly increases by a factor of \( \sim 3.8 \).

Simultaneously, the UV ionization background has also evolved significantly with redshift \citep{haardt2012}, contributing to the ionization of neutral carbon to \civ~\citep{songaila01}. The energy required for the transition \( C^{2+} \rightarrow C^{3+} \) is \( \Delta E \approx 47.9\, \rm eV \), which is very close to \( \Delta E (\text{He}^{+} \rightarrow \text{He}^{2+}) \approx 54.1\, \rm eV \). As described in \citet{haardt2012}, the optically thin photoheating rates per ion for He\textsc{ii} (see Figure 8 in \citealt{haardt2012}) have increased by two orders of magnitude from \( z \approx 5 \rightarrow 1.7 \). In addition, the mean IGM metallicity generally increases smoothly over these epochs, as shown in the previous section. The metallicity measurements do depend on the \( dN/dX \) and \( \Omega_{\rm CIV} \), suggesting a possible connection between the cosmic mass density of \civ~, IGM metallicity and the UV background evolution at these redshifts. Additionally, previous studies have found that lower-ionization carbon (C~\textsc{iii}) is less abundant than higher-ionization \civ~from \( z = 4.5 \) to \( z \sim 2.3 \) \citep{simcoe11}, which qualitatively aligns with the increase in \( dN/dX \) observed at similar redshifts in our study.  

\subsection{Possible Origin of \civ~absorbers}\label{origin}

The physical origin of these weak ($EW_{\rm 1548}<0.6\,\rm \AA$) and strong ($EW_{\rm 1548}>1\,\rm \AA$) ~\civ~absorber systems remains an open question. Cross-correlation studies of metal absorbers with foreground galaxies suggest that strong \civ~absorbers are typically associated with galactic winds, with their evolution linked to feedback processes \citep{fox2005}. As shown in \citet{hasan20}, AGN-driven winds may not be the primary driver of the redshift evolution of weak \civ~systems (\(\mathrm{EW}_{\mathrm{CIV}} \geq 0.05\)~\AA). Unlike AGN activity, which peaks around \( z \sim 2 \) \citep{kulkarni2019}, the comoving path density (\( dN/dX \)) of weak \civ~absorbers does not exhibit similar behavior and instead peaks around \( z \sim 1.4 \). This suggests that AGNs may not be solely responsible for \civ~enrichment in the IGM. On the other hand, as shown by \citet{segers2017} using EAGLE simulations, fluctuating AGNs can create `AGN proximity zone fossils’ in the CGM of galaxies, where \civ~can remain ionized for several million years after the AGN has turned off, resulting in a non-zero covering factor within \( 2R_{\rm vir} \).

The \civ~redshift evolution model developed first in \citet{hasan20} and then adapted in \citet{hasan21}, predicts that the weakest \civ~systems (\( EW_{\rm 1548} \sim 0.05 \)~\AA) should be present as early as \( z \sim 8 \) (i.e., just \( \sim 600 \) Myr after the Big Bang). In contrast, strong absorbers (\( EW_{\rm 1548} \sim 0.6 \)~\AA) do not appear until \( z \sim 4 \), when the Universe was approximately $1.8$ Gyr old. This suggests a connection between the production of strong \civ~systems and the epoch of H\,\textsc{i} reionization.

Since the cosmic reionization epoch is still not well constrained, determining when the first \civ~systems formed remains a challenging question. Recent studies have shown that galaxy formation and evolution simulations underproduce the column density distribution of \civ~systems at \( z > 5 \), despite successfully reproducing the observed history of Ly$\alpha$ absorption and Si\,\textsc{iv} \citep{finaltor20}. Simulations have revealed that \civ~bubbles may have sizes of several Mpc \citep{oppenheimer08,finaltor20} at $z\sim 10$, encompassing large comoving volumes. Given the sensitivity of JWST, it is predicted to identify \( \gtrsim10 \) galaxies \citep{bunker2023} per \civ~system at these epochs within these comoving volumes. Recent observations of four high-redshift quasar sightlines (\( z > 6.5 \)) have revealed that the incidence rate of high-ionization metal species such as \civ~and Si\,\textsc{iv} decreases at \( z \approx 6 \), while low-ionization species continue to increase \citep{christensen2023}.  

Conversely, emission-line studies have detected several galaxies and AGNs at \( z \sim 9-11 \) with JWST, showing unresolved \civ~doublet emission peaks \citep{napolitano2025}. Thus, the redshift range accessible with optical data provides only a partial view of the overall evolution of \civ~systems. A comprehensive multi-wavelength study is necessary to construct a complete picture of \civ~absorbers across a wide range of redshifts.

At the same time, studying the evolution of \civ~in large-box simulations provides a complementary approach. \citet{hasan20} compared their Keck results with the Technicolor Dawn (TD) simulation \citep{finaltor20} and found that it best reproduces the equivalent width frequency distribution for weak to intermediate-strength absorbers (\( 0.4\, \rm \AA < EW_{\rm 1548} < 0.8\, \rm \AA \)). At the weaker end, TD overproduces the incidence rates by a factor of \( \sim 2 \), while at the stronger end, it underproduces them by a factor of \( \sim 4 \). Similarly, the \textsc{Illustris} simulation \citep{nelson2015} also fails to reproduce observations \citep{bird2016} accurately. TD simulations show a better match at \( z < 5 \) than at \( z > 5 \). One possible explanation is the size and resolution limitations of the simulation, which may exclude some massive galaxies. Another factor could be the modeling of the UV background, stellar feedback, and quasar number density, all of which contribute to the abundance of \civ~absorbers at these epochs. 

Furthermore, \citet{oppenheimer06} used hydrodynamical simulations to investigate the origin of \civ~systems in the IGM. They found that momentum-driven wind models best match the observed global \civ~mass densities at \( z \sim 2-5 \). Their detailed ionization modeling also revealed that the ratio of \civ~to total carbon remains relatively independent of metallicity at these redshifts and that \civ~absorbers are primarily found in the diffuse medium. 

Additionally, the connection between absorbers and halos has revealed non-zero covering factors—the probability of detecting \civ~absorbers along a line of sight—in the outskirts of galaxies, linking them to galactic outflows \citep{tumlinson17, schroetter21}. This is consistent with the model proposed by \citet{hasan21}, which suggests that \civ~systems in the outer halos of galaxies are predominantly produced by outflows.

Overall, fully disentangling the dominant physical processes driving the evolution of \civ~absorbers remains a complex challenge, and a combination of high and low-resolution data would provide a holistic picture of \civ~evolution at a broad range of redshifts.

\section{Summary}\label{summary}

In this paper, we have compiled the largest sample of \civ~absorber systems detected in low-resolution quasar spectra using an automated doublet-finder method. This is the first \civ~catalog based on DESI quasar data from DR1 and DR2 and is \( \sim 3-6 \) times larger than the previous SDSS DR7/DR12 quasar-based catalogs \citep{cooksey10, cooksey13}.  The detection algorithm and the associated Python script are publicly available to the community. Below is a summary of our main findings:

\begin{itemize}
    \item We compile the largest and most up-to-date \civ~absorber catalog based on DESI quasars. We detect \( 101,487 \) systems with \( \mathrm{EW}_{\mathrm{CIV}} \geq 0.1 \)~\AA\ in the redshift range \( 1.3 < z < 4.5 \) (with a mean redshift of \( \braket{z} \approx 2.26 \)) \( 300,637 \) quasar sightlines in the DESI Data Release 2, a $\sim 33\%$ detection rate. The \civ~doublet properties are well within the theoretical limits, indicating high purity. 

    \item Using a Monte Carlo approach, we measure completeness as a function of absorber and quasar properties by simulating millions of mock absorbers. Our catalog is \(\sim50\% \) complete at \(EW_{\rm 1548} \sim 0.4 \)~\AA~ and reaches \( \gtrsim 95\% \) completeness at \( EW_{\rm 1548} \sim 1.4 \)~\AA. 
    
    \item We also construct a high signal-to-noise composite spectrum that shows several weak absorption lines. By comparing the average equivalent width of detected absorbers with those measured in the stacked spectrum, we estimate a catalog purity of \( \simeq 95\% \).

    \item The number of absorbers per unit rest-frame equivalent width per unit redshift path density, \(\frac{d^2N}{dW dX}\) (i.e., the absorber incidence rate), declines smoothly with equivalent width at all redshifts, indicating that weak absorbers are more prevalent than strong ones across all epochs. The behavior is well described by a simple exponential function at all redshifts.

    \item Our analysis also reveals that the redshift path density (\( dN/dX \)) follows a similar smooth trend, increasing by a factor of \( \sim 2-5 \) from \( z \approx 4.5 \rightarrow 1.4 \). Weak absorbers exhibit less evolution compared to strong absorbers.

    \item Using the apparent optical depth method (AODM), we measure the column densities of our absorber systems and estimate the \civ~cosmic mass density (\( \Omega_{\rm CIV} \)) as a function of redshift from \( z \approx 4.5 \) to \( z \approx 1.4 \). \( \Omega_{\rm CIV} \) exhibits a steep increase ($\sim 3.8$ times) as the Universe evolves. We shed light on the origin and evolution of \civ~absorbers in the context of UV background, cosmic star formation history, and He~\textsc{ii} photoheating rate.

    \item Using a simple linear relation between \( \Omega_{\rm CIV} \) and the metallicity traced by carbon, we constrain the IGM metallicity to \( \log(Z/Z_{\odot}) \gtrsim -3.25 \) at \( z \sim 2.3 \). We also find that it increases smoothly as the Universe evolves over time.
    
    \item Finally, we also discuss the future avenues where the catalog could shed light on the multiphase nature of the circumgalactic medium using absorber-galaxy correlation using large spectroscopic surveys.

\end{itemize}

\section*{Data Availability}

The \civ~catalog based on DESI DR1 quasars is publicly available as an official DESI Value Added Catalog (VAC) at \url{https://data.desi.lbl.gov/doc/releases/dr1/vac/civ-absorber/}. Additionally, we also make it available along with example notebooks on GitHub at \url{https://github.com/abhi0395/desi-dr1-civ}. The DR2 catalog (used for analysis in this paper) will be released as part of DESI DR2 VAC in the near future. The data associated with the plots in the results section are available at \url{https://doi.org/10.5281/zenodo.15061748}. The code to generate the catalog is publicly available at \url{https://github.com/abhi0395/qsoabsfind} and \url{https://doi.org/10.5281/zenodo.15685771}.

\section{Acknowledgement}
\modify{We thank the anonymous referee for providing constructive feedback that significantly improved the quality and clarity of the paper.} All the computations were performed at the DOE's high-performance computing facility, NERSC, located at Berkeley Lab. This material is based upon work supported by the U.S. Department of Energy (DOE), Office of Science, Office of High-Energy Physics, under Contract No. DE–AC02–05CH11231, and by the National Energy Research Scientific Computing Center, a DOE Office of Science User Facility under the same contract. Additional support for DESI was provided by the U.S. National Science Foundation (NSF), Division of Astronomical Sciences under Contract No. AST-0950945 to the NSF’s National Optical-Infrared Astronomy Research Laboratory; the Science and Technology Facilities Council of the United Kingdom; the Gordon and Betty Moore Foundation; the Heising-Simons Foundation; the French Alternative Energies and Atomic Energy Commission (CEA); the National Council of Humanities, Science and Technology of Mexico (CONAHCYT); the Ministry of Science and Innovation of Spain (MICINN), and by the DESI Member Institutions: \url{https://www.desi.lbl.gov/collaborating-institutions}. Any opinions, findings, and conclusions or recommendations expressed in this material are those of the author(s) and do not necessarily reflect the views of the U. S. National Science Foundation, the U. S. Department of Energy, or any of the listed funding agencies.

The authors are honored to be permitted to conduct scientific research on Iolkam Du’ag (Kitt Peak), a mountain with particular significance to the Tohono O’odham Nation. 

\software{\texttt{Matplotlib}
  \citep{hunter07a}, \texttt{NumPy} \citep{harris20a}, \texttt{Scipy} \citep{virtanen20a}, \texttt{Astropy} \citep{astropy18}, \texttt{qsoabsfind} \citep{anand2021, qsoabsfind2025}, \texttt{numba} \citep{numba2015}}

\bibliographystyle{aasjournal}
\bibliography{refs}

\appendix
\section{Composite Residual in Quasar-Frame}\label{qso_residual}
\begin{figure*}
    \centering \includegraphics[width=0.95\linewidth]{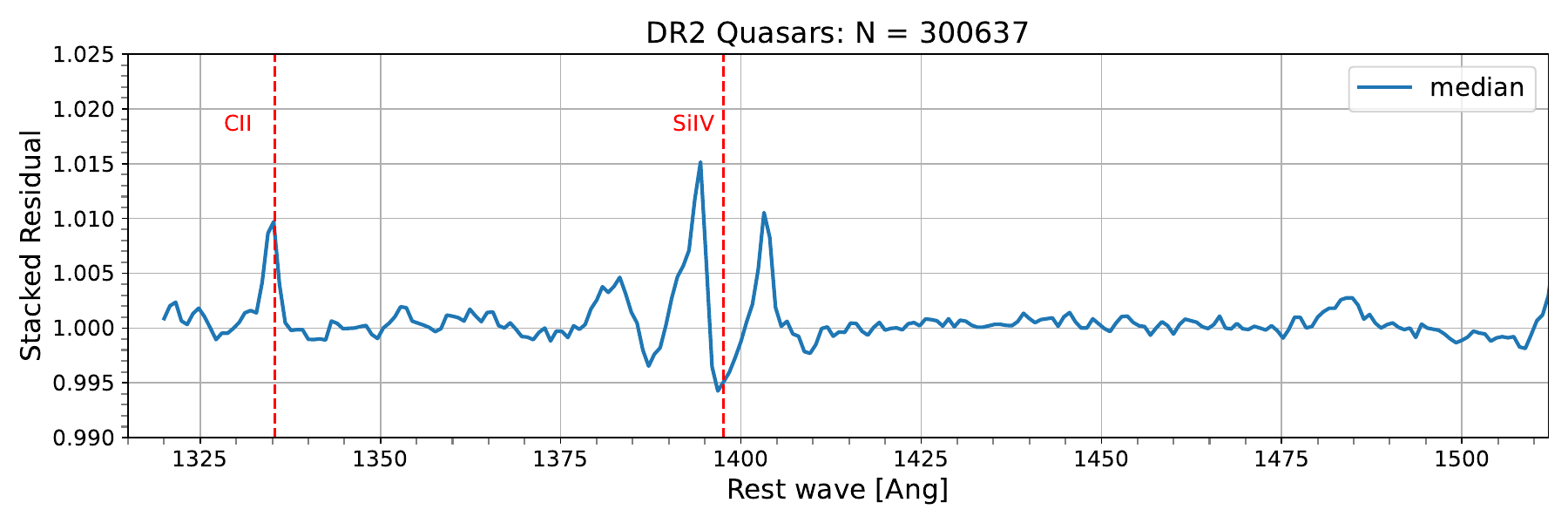}
    \caption{Median composite spectrum of residuals in the rest-frame of parent quasars. The continuum normalization yields a very flat residual with $\lesssim1-1.5\%$ variation. The wiggly features found near C~\textsc{ii} and Si~\textsc{iv} emission regions of the quasars clearly indicate that our continuum modeling does not work very well around them.}
    \label{fig:residual_qso_frame}
\end{figure*}

To assess the robustness of our continuum modeling, we constructed a composite median residual spectrum in the rest frame of the parent quasars. The composite spectrum is shown in Figure~\ref{fig:residual_qso_frame} as a function of the quasar rest-frame wavelength. We find that the residual remains mostly flat around \( R \sim 1 \), with small variations (\( \lesssim 1\% \)) near the C~\textsc{ii} and Si~\textsc{iv} emission features of quasars. This could be attributed to our continuum model, which combines mean transmission and a quasar color term, potentially underestimating the mean flux around emission features \citep{dumasbourboux2020}.  

\begin{figure}
    \centering
    \includegraphics[width=0.75\linewidth]{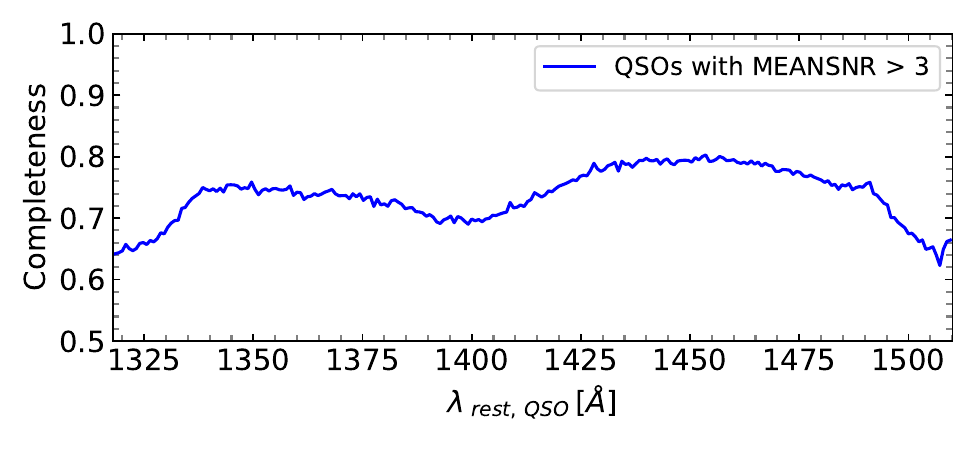}
    \caption{Average completeness of \civ~detection as a function of the rest-frame wavelength of parent quasars. As expected, completeness is lower at the edges due to increased noise. Overall, it remains fairly constant between \( 70-80\% \) within the absorber search window.}
    \label{fig:qso_completeness}
\end{figure}

As described in Section~\ref{completeness}, we also measured completeness as a function of quasar rest-frame wavelength. The results are shown in Figure~\ref{fig:qso_completeness}. We observe that completeness varies between \( \sim 70-80\% \) within the absorber search window, with slightly lower values at the edges, because we slightly offset from the edges in our absorber search window. This behavior indicates that our continuum estimation is generally reliable and does not significantly impact the completeness analysis. Ideally, completeness would be flat if the continuum models were near-perfect.

\section{Method Validation on Previous Catalog}\label{cooksey_catalog}

\begin{figure}
    \centering
    \includegraphics[width=0.7\linewidth]{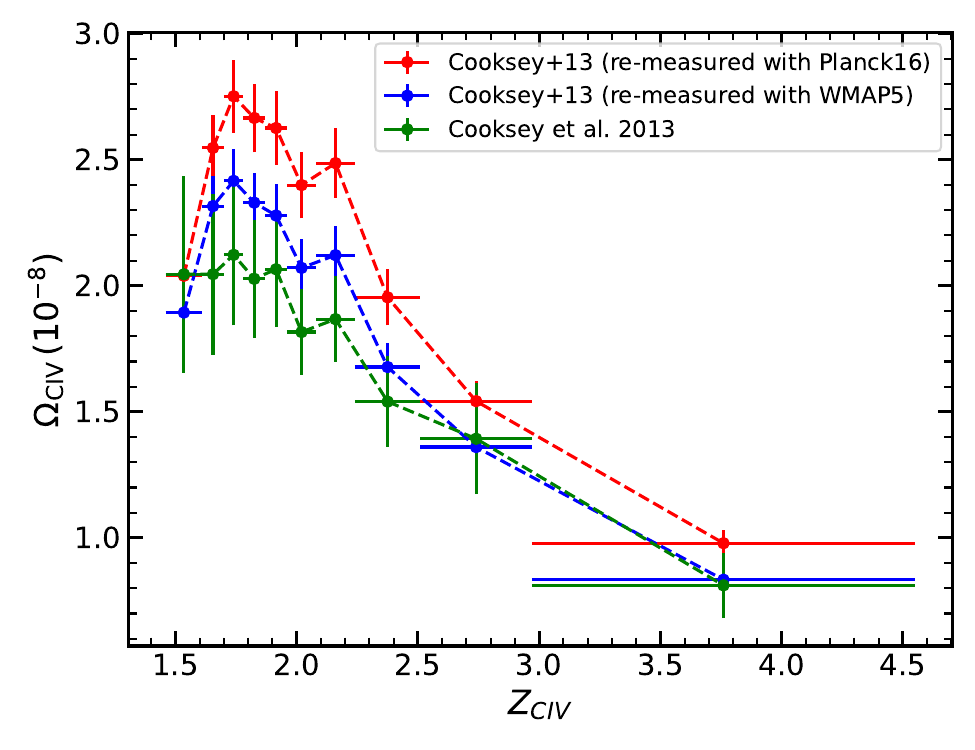}
    \caption{Remeasured the \civ~cosmic mass density using the absorber catalog of \citet{cooksey13} for different cosmologies. Our method successfully reproduces their results, further validating the accuracy of our approach.}
    \label{fig:cooksey_omega}
\end{figure}

We validated our method for measuring path densities and the cosmic mass density of \civ~absorbers using the publicly available catalog from \citet{cooksey13}. Using their \civ~redshifts, quasar catalog, equivalent widths, and completeness tables, we measured the cosmic mass density of \civ~absorbers for systems with \civewGr{0.6} detected in SDSS DR7. We estimated mass density for both Planck \citep{planck16} and WMAP5 \citep{komatsu2009} cosmologies and compared our results with their published results. The \civ~mass density as a function of redshift is shown in Figure~\ref{fig:cooksey_omega}. We recover the densities pretty well for both cosmologies, though our values are slightly higher (but within error bars), while the overall trend remains consistent. This further confirms the reliability of our method and demonstrates that it can successfully reproduce previous results.

\end{document}